\documentclass[12pt]{article}

\usepackage{amsmath}
\usepackage{amsfonts}
\usepackage{amssymb}
\usepackage{natbib}
\usepackage{graphicx}
\usepackage{pdfpages}
\usepackage[onehalfspacing]{setspace}
\usepackage[margin=1.5in]{geometry}
\usepackage{subcaption}
\usepackage{setspace}
\usepackage{placeins}
\usepackage{url}
\usepackage{xcolor}
\usepackage{mathtools}
\usepackage{titlesec}

\titlelabel{\thetitle.\quad}

\begin{document}

\def\spacingset#1{\renewcommand{\baselinestretch}%
	{#1}\small\normalsize} \spacingset{1}

	\pagenumbering{arabic}

	\title{\bf	A Dynamic Separable Network Model with Actor Heterogeneity: An Application to  Global Weapons Transfers}
	\author{Michael Lebacher\thanks{Department of Statistics, Ludwig-Maximilians-Universit\"at M\"unchen, 80539  Munich,  Germany, michael.lebacher@stat.uni-muenchen.de}\hspace{.2cm}, 
		Paul W. Thurner\thanks{Department of Political Science, Ludwig-Maximilians-Universit\"at M\"unchen, 80538  Munich,  Germany, paul.thurner@gsi.uni-muenchen.de}\hspace{.2cm} and 
 G\"oran Kauermann\thanks{Department of Statistics, Ludwig-Maximilians-Universit\"at M\"unchen, 80539  Munich,  Germany, goeran.kauermann@stat.uni-muenchen.de}\hspace{.2cm}\thanks{The authors gratefully acknowledge funding provided by the german research foundation (DFG) for the project \textit{International Trade of Arms: A Network Approach}.}}
\date{}
	\maketitle

\bigskip
\begin{abstract}
\noindent	In this paper we propose to extend the separable temporal exponential random graph
model (STERGM) to account for time-varying network- and actor-specific effects. Our
application case is the network of international major conventional weapons transfers, based on data from the Stockholm International Peace Research Institute (SIPRI). The application is particularly suitable since it allows to distinguish the potentially differing driving forces for creating new trade relationships and for the endurance of existing ones. 
In accordance with political economy models we expect security- and network-related covariates to be most important for the formation of transfers, whereas repeated transfers should prevalently be determined by the receivers' market size and military spending. Our proposed modelling approach corroborates the hypothesis and quantifies the corresponding effects. Additionally, we subject the time-varying heterogeneity effects to a functional
principal component analysis. This serves as exploratory tool and allows to identify countries that stand out by exceptional increases or decreases of their tendency to import and export weapons.
\end{abstract}

\noindent%
\vfill

\newpage
\spacingset{1} 

\section{Introduction}

In this paper we present a data-driven extension of the separable temporal exponential random graph model (STERGM, \citealt{krivitsky2014}) applied  appropriately to a highly relevant case: The international weapons exchange. 
The STERGM  allows to differentiate between the \textsl{formation}, i.e.\ new arms trades, and the \textsl{persistence} of existing edges, i.e.\ continued arms transfers. To introduce into the field, we first sketch and motivate network analysis for (arms) trade data. We then put the model in a broader context of statistical network models, supplemented by a description and discussion of international arms trade.

\noindent	\textsl{Trade networks}

\noindent Statistical network analysis provides a good framework to conceptualize  international trade systems. \citet{schweitzer2009}  highlight the enormous interdependencies of economic transactions and propose a network approach for capturing the systemic  complexity. Gravity models, as standard approach in econometrics for modelling trade data (\citealt{head2013}), are usually focussed on dyadic relations. Hence, the models exclude highly important hyper-dyadic dependencies, and especially indirect relations. \citet{squartini2011a,squartini2011b} showed that gravity models of international trade are, therefore, necessarily incomplete. In particular, they demonstrated that analysing the determinants of link creation is highly important as the binary network carries information that goes beyond the classical gravity model representation. \citet{barigozzi2010} demonstrated that trade networks are commodity-specific, i.e.\ their topologies are quite different across commodities - leading us to conclude that  there is also a need to consider arms transfers separately. This is theoretically challenging since arms transfers constitute a very special trade relationship. The transferred products and services can potentially lead to deadly quarrels between or within states, or they may contribute to stabilization and deterrence. The delivery is not always a purely economic exchange but may also serve the support of aligned countries or groups. In sum, the exchange of weapons is a politically sensible and security-related, but also an economically beneficial relationship. For this reason, we make use of flexible statistical models for network data that allow us to investigate the special incentives in the international arms trade network.

\noindent	\textsl{Statistical network models}

\noindent	Statistical models that are suitable for temporal networks have been developed just in the recent years, and different techniques have been proposed. \citet{robins2001} were the first to extend the static exponential random graph model (ERGM, \citealt{holland1981}; \citealt{lusher2012})
to discrete-time Markov chain models, see also \citet{snijders2010}. \citet{hanneke2010} or \citet{leifeld2017} also consider network dynamics on a discrete time scale. They propose the  temporal exponential random graph model (TERGM) which makes use of a Markov structure conditioning on previous network statistics as covariates in the model. A related approach is presented by \citet{almquist2014}, discussing assumptions that allow for circumventing the often computationally intractable fitting process of dynamic network models by applying logistic regression models. \citet{koskinen2015} expand the model using Bayesian methods which allows the parameters in the dynamic network model to  change with time. A general perspective on dynamic networks is provided by \citet{holme2015}. It also includes models for continuous time, such as stochastic actor-oriented models (SAOM, see \citealt{snijders2010soam}) or dynamic stochastic block models (SBM, see for instance \citealt{xu2015}).

A recent novel modelling strategy for networks observed at discrete time points has been proposed by \citet{krivitsky2014}. They do not model the state of the network itself but rather focus on network changes which either occur because of the formation of new edges  or because of the (non-)persistence of existing ones. Assuming independence between the two processes, conditional on the previous network, leads to the so called \textsl{separable} TERGM. The separation is motivated by the fact that the two processes under study are highly likely to be driven by different mechanisms and factors. The authors argue that the inclusion of a stability term (being mathematically equivalent to the inclusion of the lagged edge values as explanatory variable) in a TERGM could lead	 to ambiguous conclusions because it is not clear whether a positive stability parameter means that non-existing ties remain non-existent (no formation) or whether existent ties remain existent (persistence).

For many real world dynamic networks the process change with time and therefore the assumption of stationarity seems to be inappropriate. This is especially the case for network data that span a long time period and potentially subject to structural breaks. Under such conditions it appears necessary to allow the model parameters to change with time. We take up this idea and extend the STERGM by allowing for time-varying coefficients. More specifically, we propose to rely on so called generalized additive models (GAM). This model class has been proposed by \citet{hastie1990} and extended fundamentally by \citet{wood2017} to allow for smooth, semi-parametric modelling of time-varying parameters in a generalized regression framework (see also \citealt{ruppert2009}).

Furthermore, the assumption of node homogeneity must be regarded as questionable. We therefore allow for heterogeneity in the model (see  \citealt{thiemichen2016} for a discussion on node heterogeneity). Accordingly, we follow the $p_2$-model developed by \citet{duijn2004}  and enrich the STERGM with functional time-varying random effects  (\citealt{durban2005}) which leads to smooth node-specific effects. We propose to investigate the fitted functional heterogeneity effects with techniques from functional data analysis (FDA), see for instance \citet{ramsay2005}. This allows to identify countries (nodes) that have fundamentally changed their role in the arms-trading network over the observation period.

\noindent	\textsl{Global weapons transfers} \label{usecase}

\noindent
At present, there are only a few empirical binary network analyses of the international arms trade.
\citet{akerman2014} pioneered in analysing topological features of the binary arms trade network. Their descriptive network analysis is supplemented by an empirical investigation using a binarized gravity model without considering network dependencies. 
In this article we build on the recently published paper by \citet{Thurner2018} that uses a TERGM. However, our approach extends the TERGM in many aspects. Most importantly, we treat dynamic dependencies in a fundamentally different way. In \citet{Thurner2018}, the authors found that previous arms trading has a highly determining impact on the occurrence of subsequent transfers due to the enormous inertia. This finding implies that the information whether trade happened in the preceding time period(s) has a considerable impact on the probability to trade again,  leading to the same ambiguities as mentioned in the stability term example by \citet{krivitsky2014}.  In order to disentangle the driving network formation forces due to pure inertia, we propose to incorporate this distinction directly in the model. More precisely,  the STERGM allows us to investigate whether the mechanisms that result in transfers being formed without immediate predecessor differ from those that lead to consecutive transfers. This is also of practical importance because governments carefully reflect the decision whether to authorize arms transfers based on economic and security considerations. Furthermore, they continuously reconsider this decision whether to maintain such trade relations or whether to dissolve because the importer potentially jeopardises strategic interests or violates once shared normative standards (see \citealt{garcia2007} for the general model and for  \citealt{Blanton2005} as well as \citealt{eric2015} for normative considerations).

We expect several necessary conditions to hold for the formation of transfers: the receiving country must be considered at least marginally trustworthy and politically and economically reliable. 
Hence, passing a threshold of trustworthiness is required for formation, i.e\ building new trades. The special role of trustworthiness in arms transfers stems from the fact that security concerns play an important role when governments decide whether to license the delivery. We expect network statistics, as well as regime dissimilarity and formal alliances to play a prominent role in the formation stage to raise a relationship above the minimum threshold level of reservation. Follow-up trades and their repetition should then be rather dominated by economic considerations like the size of a receiver economy and by the size of the military expenditures (see \citealt{Schulze2017}).

While differentiation between formation and repetition, respectively, legitimates the use of the STERGM per se, our extensions of the model towards time-varying coefficients are important and in our view inevitable because the observational time covers more than 65 years. Hence, the introduction of smooth dynamic effects is needed to build a realistic model. Given the dynamic evolution of the network, the historical developments and the presence of  at least one system-wide structural break with the collapse of the Soviet Union, we expect that the generative mechanisms change over time and differ with respect to the included variables if we compare the pre- and post cold war time period (see also \citealt{akerman2014} and \citealt{Thurner2018}).

Finally, we argue that not all network activities and trades can be explained by observables and, thus, unobserved heterogeneity remains. We expect primarily actor-specific heterogeneity which is accentuated by systematic historical accounts (\citealt{Harkavy1975}; \citealt{krause1995arms}). This  highlights the self-reinforcing tendencies of technological advantages of highly developed countries which results in strong heterogeneity of the countries' abilities to export (and import).  Therefore, the inclusion of actor-specific random effects seems necessary and we expect strong heterogeneity among the countries with respect to imports and exports.

\vspace{0.1cm}

We proceed as follows. Section
\ref{descrip} presents the data provided by the Stockholm International Peace Research Institute (SIPRI). Section \ref{model} introduces the statistical models used to analyse the data. Section \ref{results} provides the results and their interpretation. Section \ref{conc} concludes the paper.

\section{Data description and preprocessing}\label{descrip}
\begin{figure}[t!]
	
	\centering
	\includegraphics[trim={0cm 0cm 0cm 0cm},clip,width=0.9\textwidth]{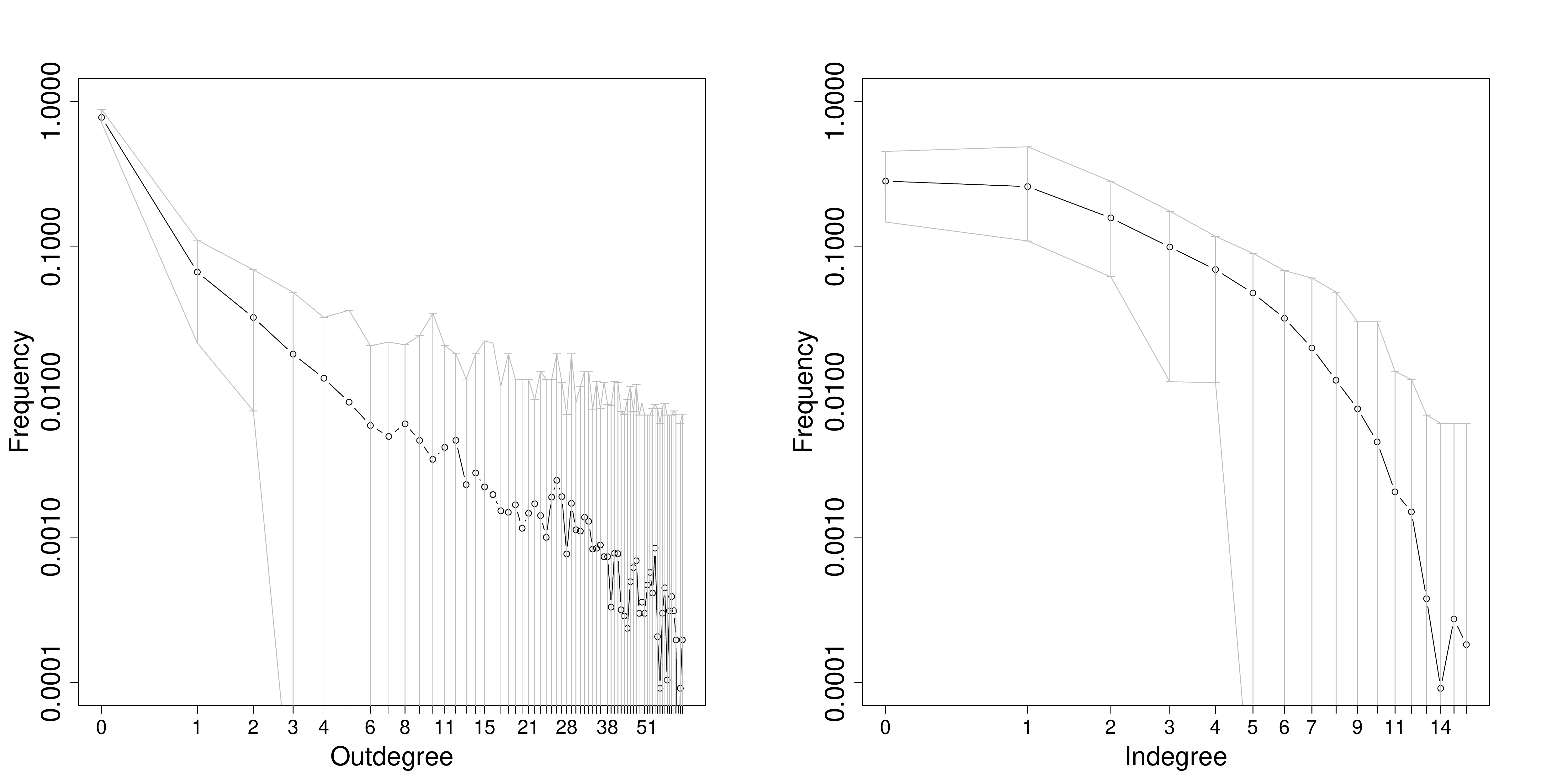}
	\caption{Degree distributions of the included countries for the outdegree (number of outgoing edges) on the left and indegree (number of ingoing edges) on the right. Averages over all years are represented by the solid line. The whiskers in grey show the minimum and maximum Values realized in all years. Both axes are in logarithmic scale.}
	\label{degree_dist}
\end{figure}
Data on the international trade of major conventional weapons (MCW) are provided by the Stockholm International Peace Research Institute (see \citealt{sipridata2017}). They include for example aircrafts, armoured vehicles and ships (see Table \ref{coverage} in the Appendix \ref{descrannex} for an overview of the types of arms).  
The countries included and their three-digit country codes are given in Table \ref{count_inc} of Appendix \ref{descrannex}. Note that we have excluded all non-state organizations like the Khmer Rouge or  the Lebanon Palestinian Rebels from the dataset as well as countries with no reliable covariate information available.

Figure \ref{Fig-network-1} in the Appendix \ref{descrannex} shows binary networks for the years 2015 and 2016 and Figure \ref{sum_stat} in the Appendix \ref{descrannex}  provides a collection of summary statistics for the networks.

We focus on the binary occurrence of trade thereby disregarding the exact  transfer volumes and follow \citet{akerman2014} and \citet{Thurner2018} in setting the edge value to one if there is a trade flow greater zero between two countries and zero else. Additionally, we re-estimated our model with different thresholds and found that the results are quite robust, for details see the Supplementary Material.
\begin{figure}[t!]
	\centering
	\includegraphics[width=0.9\textwidth]{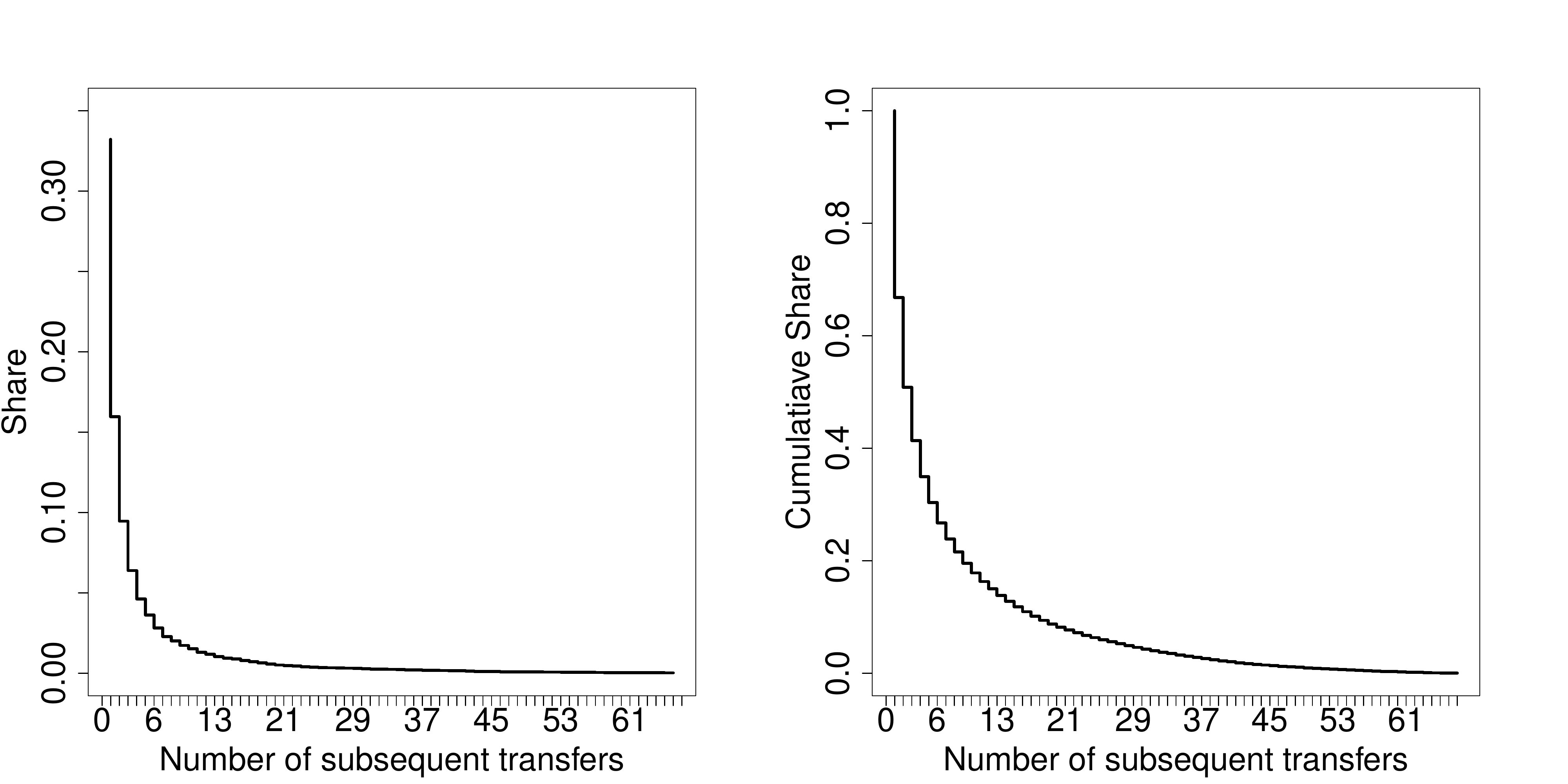}
	\caption{Share of subsequent arms transfers (left) and cumulative share of subsequent arms transfers (right). Number of subsequent transfers on the  horizontal axis and share of observations on the vertical axis.}
	\label{fig:d1}
\end{figure}

The analysis of the degree distributions is of vital interest in statistical network analysis (\citealt{barabasi1999}) and gives important insights into the basic properties of the network under study. With more than 65 networks to analyse, we compute the period-average degree distribution and provide information on the minimal and maximal value of the realized degree distribution. This is represented in a log-log version in Figure \ref{degree_dist} for both, outdegree and indegree. The plot shows the enormous heterogeneity in the networks. Most of the countries have no exports at all with a time-average share of 78\% of countries exhibiting outdegree zero, while the outdegree distribution has a long tail, indicating that there are a few countries, having a very high outdegree. The highest observed outdegree in a year is 66 and is observed for the United States. Other countries with exceptional high outdegree for almost the whole time period are Russia (Soviet Union), France, Germany, United Kingdom, China, Italy and Canada. In the right plot, the indegree distribution can be seen. Here the pattern is different. The highest value observed in a year is 16 and corresponds to Saudi Arabia. In contrast to the outdegree distribution, the countries with a high indegree are changing with time. In the beginning of the observational period the countries with the highest indegree were Germany, Indonesia, Italy, Turkey and Australia, but in more recent times these are the United Arab Emirates, Saudi Arabia, Singapore, Thailand and Oman.

In Figure \ref{fig:d1} we provide a graphical representation of the change and stability patterns in the network. On the left hand side we present the share of observations (vertical axis) against the number of subsequent transfers (i.e.\ repeated transfers)  on the horizontal axis. Out of roughly $19,000$ recorded trading instances only $33\%$ do not have at least one consecutive transfer in the follow-up year of a trade. Looking on the right hand side of Figure \ref{fig:d1} we visualize the share of observations (vertical axis) that has at least as much subsequent transfers as indicated by the horizontal axis. It can be seen that roughly the same share of observations ($35\%$) lasts at least five periods and almost $10\%$ of all dyadic relations last more than 20 consecutive years without any interruption. Therefore, a differentiated approach to the explanation of formation and persistence could be fruitful in this application case.  
\FloatBarrier
\section{Model}\label{model}
\subsection{Dynamic formation and Persistence model}
In this section we formalize our network model.
Let $Y^t$ be the network at time point $t$,  which consists of a set of actors, labelled as $A^t$ and a set of directed edges, represented through the index set $E^t=\{(i,j):i,j\in A^t\}$. Note that this is a slight misuse of index notation since $Y^t_{ij}$ does not necessarily refer to the $(i,j)$-th element if we consider  $Y^t$ as adjacency matrix. This is because the actor set $A^t$ is allowed to change with time, so that $i$ and $j$  are not running indices from $1$ to $n_t$, where $n_t$ is the  number of elements in $A^t$. Instead indices $i$ and $j$ represent the $i$-th and $j$-th country, respectively. We define $Y^t_{ij}=1$ if country $i$ exports weapons to country $j$ and since self-loops are meaningless, elements $Y^t_{ii}$ are not defined.

We aim to model the network in $t$ based on the previous  year network in  $t-1$. To do so we have to take into account that the actor sets $A^{t-1}$ and $A^t$ may differ. In particular we have to consider the case of newly formed countries. New countries of interest are those that are present in $t$ but do not provide information about their network embedding in the previous period. For exports this is not a concern as it is almost never the case that a new country starts sending arms immediately after entering the network. Notable exceptions are Russia, the Czech Republic and Slovakia. However, these countries have clear defined predecessor states (the Soviet Union and Czechoslovakia) which can be used in order to gain information about the position of these countries in the precedent network. Regarding the imports, there is a share of countries that start receiving arms immediately with entering the network. Notwithstanding, those transactions  represent a share of less then 0.3\% of the observed trade flows. Therefore, we regard this cases as negligible and include in the model only countries where information on the current and previous time period is available. We formalize this approach by defining $Y^{t,t-1}$ as the subgraph of $Y^t$ with actor set $B^{t,t-1}=A^{t}\cap A^{t-1}$ containing $n_{t,t-1} \coloneqq |B^{t,t-1}|$ elements. Accordingly, $Y^{t-1,t}$ represents the subgraph of $Y^{t-1}$ with actor set $B^{t,t-1}$. Note that both subgraphs share the same set of actors and $Y^{t-1}=Y^{t-1,t}$ if $A^{t-1}$ and $A^{t}$ coincide.

From a modelling perspective, we follow \citet{hanneke2010} and assume that the network in $t$ can be modelled given preceding networks, using a first-order Markov structure to describe transition dynamics for those actors included in the set $B^{t,t-1}$. Furthermore, we want to identify the driving forces of a transfer in $t$ if there was a preceding transfer in $t-1$ in the persistence model while the formation model considers the process of forming a trade relationship without a preceding transfer, i.e.\ biannual data. The notion of formation and persistence can be amended by using broader time windows. We demonstrate the robustness of our results with respect to broader time windows in the Supplementary Material.

Let $Y^{+}=Y^{t,t-1}\cup Y^{t-1,t}$ represent the formation network, that consists of edges that are either present in $t$ \textit{or} in $t-1$. For the persistence network, we define $Y^{-}=Y^{t,t-1}\cap Y^{t-1,t}$, being the network that consists of edges that are present in $t$ \textit{and} in $t-1$. Based on the actor set $B^{t,t-1}$ and given the formation and  persistence network as well as the network in $t-1$ the network in $t$ is uniquely defined by
\begin{equation}
\label{construct}
Y^{t,t-1}=Y^{+}\backslash ( Y^{t-1,t}\backslash Y^{-}) =Y^{-} \cup (Y^{+} \backslash Y^{t-1,t}).
\end{equation}
Note that both, $Y^+$ as well as $Y^-$ depend on time $t$ as well, which we omitted in the notation for ease of readability.
We assume that for each discrete time step, the processes of formation and persistence are separable. That is, the process that drives the formation of edges does not interact with the process of the persistence of the edges conditional on the previous network. Formally this is given by the conditional independence of $Y^+$ and $Y^-$:
\begin{equation*}
\begin{split}
&P(Y^{t,t-1}=y^{t,t-1}|Y^{t-1,t}=y^{t-1,t};\theta)=\\& P(Y^{+}=y^{+}|Y^{t-1,t}=y^{t-1,t};\theta^+)P(Y^{-}=y^{-}|Y^{t-1,t}=y^{t-1,t};\theta^-),
\end{split}
\end{equation*}
where the lower case letters denote the realizations of the random networks and $\theta=(\theta^+,\theta^-)$ gives the parameters of the model. We will also include non-network related covariates in our analysis, but we suppress this here in the notation for simplicity.

Note that it is not possible to use the lagged response as predictor, as by construction $Y^{t-1,t}_{ij}=1 \Rightarrow Y^{+}_{ij}=1  $ and $	  Y^{t-1,t}_{ij}=0\Rightarrow Y^{-}_{ij}=0   $. That is, an edge that existed in $t-1$ cannot be formed newly and an edge that was not existent in $t-1$ cannot be dissolved. It follows that the formation model  exclusively focuses on the binary variables $Y_{ij}^{+}$ with $(i,j) \in E^+= \{ (i,j): i,j \in B^{t,t-1}, Y^{t-1,t}_{ij}=0 \}$. This assures that in $t-1$ no edge between actors $i$ and $j$ was present and both actors are observable at both time points. Equivalently, the model for $Y^-$ consists of observations $Y_{ij}^-$ with  $(i,j) \in E^-= \{ (i,j): i,j \in B^{t,t-1}, Y^{t-1,t}_{ij}=1 \}$, assuring that only edges that could potentially persist enter the model. The time-dependence of $E^+$ and $E^-$ is omitted for ease of readability.

If we use an ERGM for the transition, this would yield the following probability model for the formation
\begin{equation*}
P(Y^+=y^+|Y^{t-1,t}=y^{t-1,t};\theta^+)=\frac{\exp\{\theta^+g(y^{+},y^{t-1,t})\}}{\sum_{\tilde{y}^+ \in \mathcal{Y}^+(y^{t-1,t})}\exp\{\theta^+g(\tilde{y}^+,y^{t-1,t})\}}.
\end{equation*}
The sum in the denominator is over all possible formation networks from  the set of potential edges that can form given the network $y^{t-1,t}$. The inner product $\theta^+g(y^{+},y^{t-1})$, relates a vector of statistics $g(\cdot)$ to the parameter vector $\theta^+$. The analogous model is assumed for the persistence of edges and not explicitly given here for the interest of space.

We will subsequently work with a simplified model which is computationally much more tractable. We assume that the formation or persistence of an edge at time point $t$ does solely depend on the past state but not on the current state of the network. This is achieved by restricting the statistics such, that they decompose to
\begin{equation*}g(y^{+},y^{t-1,t})=\sum_{(i,j) \in E^{+}}y^+_{ij}\tilde{g}_{ij}(y^{t-1,t}) 
\end{equation*}
for some statistics $\tilde{g}(\cdot)$. 
This assumption is extensively discussed by \citet{almquist2014} and can be well justified by the notion that the lagged network accounts for the major share of the dependency among the edges in the current network. It also allows for intuitive interpretations as can be seen as follows. Let $Y^+_{-ij}$ represent the formation network $Y^+$, excluding the entry $Y^+_{ij}$. Then, for $(i,j)\in E^+$ the following logistic model holds
\begin{equation}
\label{logit}
\begin{split}
\log\bigg{\{} \frac{P(Y^+_{ij}=1|Y^+_{-ij}=y^+_{-ij},Y^{t-1,t}=y^{t-1,t};\theta^+)}{P(Y^+_{ij}=0|Y^+_{-ij}=y^+_{-ij},Y^{t-1,t}=y^{t-1,t};\theta^+)} \bigg{\}} &=\log\bigg{\{}  \frac{P(Y^+_{ij}=1|Y^{t-1,t}=y^{t-1,t};\theta^+)}{P(Y^+_{ij}=0|Y^{t-1,t}=y^{t-1,t};\theta^+)}\bigg{\}}\\&=\theta^+\tilde{g}_{ij}(y^{t-1,t}).
\end{split}
\end{equation}
Note that model (\ref{logit}) describes network dynamics, but the model itself is static. Hence we model dynamics but do not allow for dynamics in the model itself. This is a very implausible restriction which we give up by allowing the model parameters to change with time $t$, that is we replace the parameter $\theta^+$ by $\theta^+(t)$, representing a smooth function in time. In other words, we allow the parameters  in the model to smoothly interact with time. This leads to a time-varying coefficient model in the style as proposed by \citet{hastie1993}. The focus of interest is therefore not only on the formation and persistence of edges (trade flows) but also on how these effects change in the 67 years long observation period.

\subsection{Network statistics and explanatory variables}
From a statistical point of view, network statistics are required in order to capture network dependencies. However, as social network literature has shown, network statistics usually are not just statistical controls but convey substantial meaning (see e.g.\  \citealt{snijders2011}).  In the given context, they can be motivated by political, strategic and economic arguments that refer to real-world processes (see \citealt{Thurner2018}).	Note, that we norm all network statistics (with the exception of {\sl Reciprocity}) to be within a percentage range between 0 and 100, this is necessary in order to make the statistics independent from the varying network size and allows to compare them.
\\
{\sl Outdegree}: The outdegree of a node is a standard statistic in network models. Formally, the outdegree of actor $i$ at time point $t-1$ is defined as
\begin{equation*}
outdeg_{t-1,i}=\frac{100}{ n_{t,t-1}-1}\sum_{k \in B^{t,t-1}} y^{t-1,t}_{ik}.
\end{equation*}
The arms trade network exhibits a highly oligopolistic structure with a few high-intensity traders, hence a positive coefficient for the outdegree of the sender ($sender.outdeg_{t-1,i}$) is plausible. However, we incorporate country-specific random effects in the model and it is therefore not clear whether the senders' outdegree as a global measure is still of relevance once controlled for the random country heterogeneity. 

Only few advanced countries within NATO export and import at the same time. They have a highly differentiated portfolio, rendering specialization economically reasonable and strategically non-hazardous.  In order to better represent this world-wide asymmetry we include the outdegree of the importer ($receiver.outdeg_{t-1,j}$). This should not be captured by the random effects and we expect a clear negative effect, indicating that strong exporters seldom match with strong importers.  
\\
{\sl Reciprocity}: 	This statistic is intended to detect whether there is a general tendency of arms transfers to be mutual. The statistic measures whether the potential receiver was a sender in the dyadic relationship in the previous period:
\begin{equation*}
recip_{t-1,ij}=y^{t-1,t}_{ji}.
\end{equation*} 
Reciprocation is an essential mechanism in human relations in general, and in trade more specifically. Similar as noted above, in the context of arms transfers, especially highly developed countries exhibit this feature. Since this group of countries is rather small, and specialization-induced transfers between developed countries do not lead to continuous inflows we expect this mechanism to be rather visible at the formation stage, whereas it should not be a dominant feature for permanent repetition.
\\
{\sl Transitivity}: 
Hyperdyadic trade relationships are an effective mechanism for pooling risks in buyer-seller networks (\citealt{NetworksinInternationalTrade}) and for the emergence of generalized trust which is especially important in exchanging security goods.
As a measure for higher-order dependencies we include transitivity, defined as 
\begin{equation*}
trans_{t-1,ij}=\frac{100}{n_{t,t-1}-2}\sum_{k \in B^{t,t-1}, k \neq i,j}y^{t-1,t}_{ik}y^{t-1,t}_{kj}.
\end{equation*} 
This statistic essentially counts the directed two-paths from $i$ to $j$ in $t-1$ and can be interpreted as a direct application of the \textsl{Friend of a Friend} logic from social networks to arms trade. Clearly, this kind of network embeddedness of weapons transfer deals  is important for establishing for new ones but is also likely to be relevant  for the continuation of already existing ones. 
\\
{\sl Shared Suppliers}:
We also include a statistic that we call shared-suppliers in this context. This statistic counts the shared number of actors that export to a given pair of countries:
\begin{equation*}
sup_{t-1,ij}=\frac{100}{n_{t,t-1}-2}\sum_{k \in B^{t,t-1}, k \neq i,j}y^{t-1,t}_{ki}y^{t-1,t}_{kj}.
\end{equation*} 
This statistic allows to investigate whether two countries that share multiple suppliers have the tendency to engage in trade with each other. Such a pattern is likely to be induced by a general hierarchy in the network (see \citealt{krause1995arms}). While the first tier consists of strong exporters, the second tier is populated by countries with the ability to produce and export that are nevertheless mainly supplied by the big exporters. Countries with many shared partners are likely to  engage in trade with each other but on the other hand they are typically dependent on imports from the first tier. Therefore, relationships among those countries are rather of a sporadic nature and unlikely to endure. Consequently, we expect a positive coefficient in the formation model and a negative one in the persistence model.

Naturally, the network of international arms trade is not exclusively driven by endogenous network processes but also influenced by variables from the realms of politics and economics. We lag all exogenous covariates by one year, first in order to be consistent with the idea that the determination of the network in $t$ is based on the preceding time period and second, to account for the time lag between the ordering and the delivery of MCW.
\\
{\sl Formal Alliance}:
We regard dyadic formal alliances (including defence agreements and non-aggression pacts) as an important security related criteria that plays a central role for the formation during the cold war period. Therefore,  the binary variable $alliance_{ij}$ is included in the model, being one if countries $i$ and $j$ had a formal alliance in the previous period. Given the restriction that the data is available only until 2012 (\citealt{Defagr2016}) we extrapolate the data, thereby assuming that the formal alliances did not change between 2012 and 2015.
\\
{\sl Regime Dissimilarity}:
Another important security related variable that potentially acts on the formation of arms transfers is given by the differences in political regimes between two potential trading partners. Hence we include the so called polity IV score,  ranging from the spectrum $-10$ (hereditary monarchy) to $+10$ (consolidated democracy). This data can be downloaded as annual cross-national time-series until 2015, see \citet{Polity2016} for the data and \citet{marshall2017} as a basic reference. In our model we operationalise the distance between political regimes by using the absolute differences between the scores: $poldiff_{ij}=|polity_i - polity_j|$. 
\\
{\sl GDP}: Following the standard gravity model, we include market sizes and distance in our model. The standard measure for market size is the gross  domestic product (GDP, in millions). We include the GDP in logarithmic form for the sender ($gdp_i$) as well as the receiver ($gdp_j$). The GDP data are taken from
\citet{gleditsch2013} and merged from the year 2010 on with recent real GDP data from the World Bank real GDP dataset (\citealt{GDP2017}). Clearly, the market size and economic reliability of the exporter is a prerequisite for forming and maintaining arms exports.
\\
{\sl Distance}:
For gravity models applied to trade in commercial goods, there exists mounting empirical evidence that distance is a relevant factor for determining trade relations (\citealt{disdier2008}). We do not expect that trade costs and geographical distance impede arms trade  because arms transfers establish world-wide alignments of exporters pursuing global strategic interests. Nevertheless, we include the logarithmic distance between capital cities in kilometres (\citealt{gleditsch2013d}) in order to fulfil the gravity model specification.
\\
{\sl Military Expenditures}:
We propose to include military expenditures of the sending and receiving country. 
This measure can be used as representing the size of the defence industrial base of the exporter, and the spending power and the intensity of the threat perceptions of the importing country. Accordingly, military expenditure  is added separately for the exporter and the importer in logarithmic form ($milex_i$, $milex_j$). With regard to the distinction between formation and persistence, our  expectation is related to the hypothesis that countries with high military expenditures are attractive customers for repeated importing. We therefore expect a positive and high coefficient for the military expenditures of the importer in the persistence model. The data are available from \citet{cinc2017} in the national material capabilities data set  with \citet{singer1972} as the basic reference on the data.
\subsection{Modelling heterogeneity}\label{heterogen}
The proposed network model assumes homogeneity, meaning that all differences between nodes in the network are fully described by the gravity model, enriched by security related criteria and network statistics as proposed above. However, the arms transfer network exhibits a  rather small number of countries that are high-intensity exporters and a large number of countries that are restricted to imports.  Furthermore, there are some countries that change their relative position in the trade network during the course of time. This mirrors a  substantial amount of dynamic heterogeneity which need to be taken into account.

This dynamic heterogeneity is accommodated by the inclusion of latent country effects, capturing the unobserved heterogeneity. We to follow the idea of \citet{durban2005} and model country specific random curves which are fitted with penalized splines. This can be written in a mixed model representation such that the smooth country-specific effects are constructed using a B-spline basis with (a-priori) normally distributed spline coefficients. We follow the modelling strategy of \citet{durban2017} and assume that the model includes two time-dependent random coefficients $\phi^+_{i,sender}(t)$ and  $\phi^+_{j,receiver}(t)$. The effects are  assumed to be a realization of a stochastic process with continuous and integrable functions. For each sender and receiver in both models the country-specific curves are given by
\begin{equation}
\label{speci2}
\phi_{i}(t)=B(t)a_i
\end{equation}
where $B(t)=(B_1(t),...,B_Q(t))$ is a B-spline basis covering the time range of observations and $a_i=(a_{i1},...,a_{iQ})$ is the coefficient vector. We impose the prior distribution
\begin{equation*}
a_i \sim N(0,\sigma^2_a D_Q)\text{, }i.i.d. \text{ for }i=1,...,n
\end{equation*}
where $D_Q$ is the inverse of a difference based penalty matrix which guarantees smoothness of the fitted curves $\phi_i(t)$ (see e.g.\ \citealt{eilers1996}, for details on smoothing with B-splines). Note that for time windows where a country did not exist, the corresponding B-spline does take value zero, so that no heterogeneity effect is present.

\subsection{Complete model and estimation}
Putting all the above elements together, the specification of the formation model of equation (\ref{logit}) is given by
\begin{equation*}
\begin{split}
\theta^+(t)\tilde{g}_{ij}(y^{t-1,t},x^{t-1,t}_{ij})=&\theta_0^+  sender.outdeg_{t-1,i}\theta_1^+(t)+ receiver.outdeg_{t-1,j}\theta_2^+(t)\\&+ recip_{t-1,ij}\theta_3^+(t) +
trans_{t-1,ij} \theta_4^+(t)+sup_{t-1,ij} \theta_5^+(t)\\
&+ distance_{t-1,ij } \theta_5^+(t)+alliance_{t-1,ij} \theta_6^+(t)+poldiff_{t-1,ij} \theta_7^+(t)\\
&+ gdp_{t-1,i } \theta_8^+(t)+gdp_{t-1,j} \theta_9^+(t)+milex_{t-1,i}\theta_{10}^+(t)+milex_{t-1,j} \theta_{11}^+(t)\\ &+ \phi^+_{i,sender}(t)+\phi^+_{j,receiver}(t).
\end{split}
\end{equation*}
Analogously we get the persistence model. Estimation is carried out with spline smoothing. That is, we replace the coefficients by 
\begin{equation*}
\theta_k (t)=B(t)u_k,
\end{equation*}
where $u_k$ is penalized through
\begin{equation*}
u_k \sim N(0,\sigma^2 D).
\end{equation*}
Like above, the penalty matrix is appropriately chosen (see e.g.\ \citealt{wood2017}) and $B(t)$ is a B-spline basis. Hence, smooth functions and smooth random heterogeneity can be estimated in a coherent framework (see \citealt{durban2005}). The entire model can be integrated in the flexible generalized additive model (GAM) framework provided by \citet{wood2017}  (see also \citealt{wood2006}) which is implemented in the \texttt{mgcv} package (version 1.8-28) by \citet{Wood2011}. The identification of the smooth components and the intercept term is ensured by a "sum-to-zero" constraint (\citealt{wood2017}). For further details see the Appendix \ref{estim_detail}.
\FloatBarrier
\section{Results}\label{results}

\FloatBarrier
\subsection{Time-varying fixed effects}
\begin{figure}[t!]
	\centering	
	\begin{subfigure}{.84\textwidth}
		\centering
		\includegraphics[width=0.94\textwidth,trim={0.1cm 0cm 2cm 0cm}]{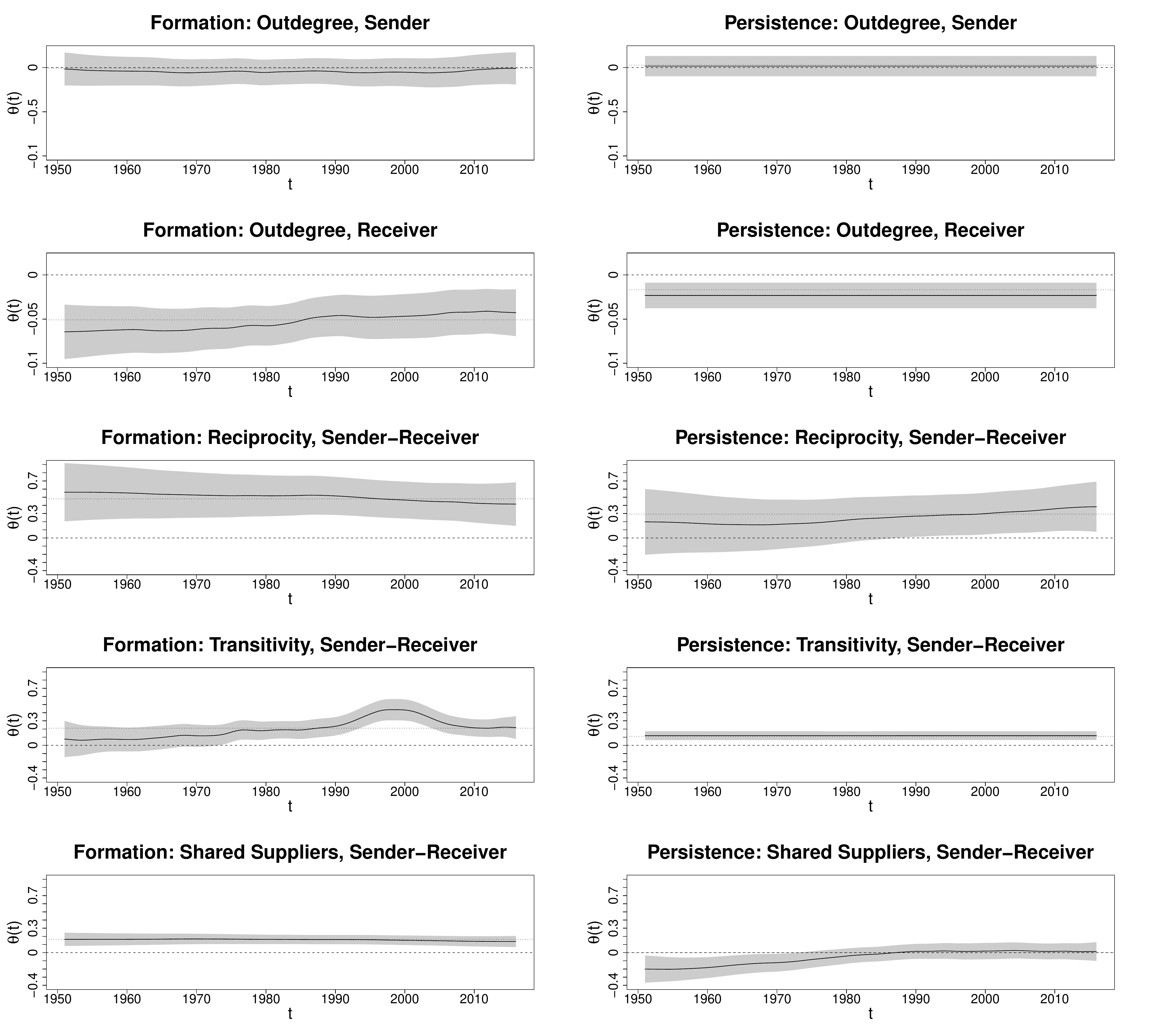}
	\end{subfigure}%
	\begin{subfigure}{.16\textwidth}
		\centering
		\includegraphics[page=1,width=1.3\textwidth,trim={0.55cm 0cm 24cm -1cm},clip]{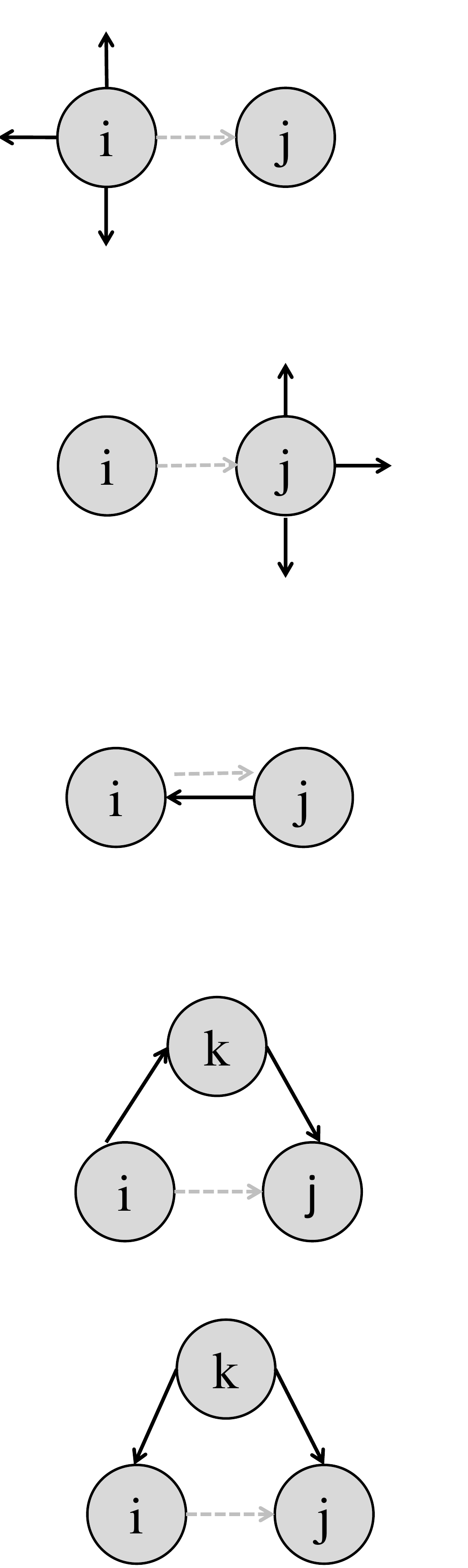}
		\label{fig:sub2}
	\end{subfigure}
	\caption{Time-varying coefficients of network statistics in solid black. Shaded areas give two standard error bounds. Time-constant effects in dashed grey and zero line in dotted black. Schematic representation of the network effects on the right hand side.}
	\label{res_net}
\end{figure}

\begin{figure}[t!]
	\centering
	\includegraphics[width=0.95\textwidth]{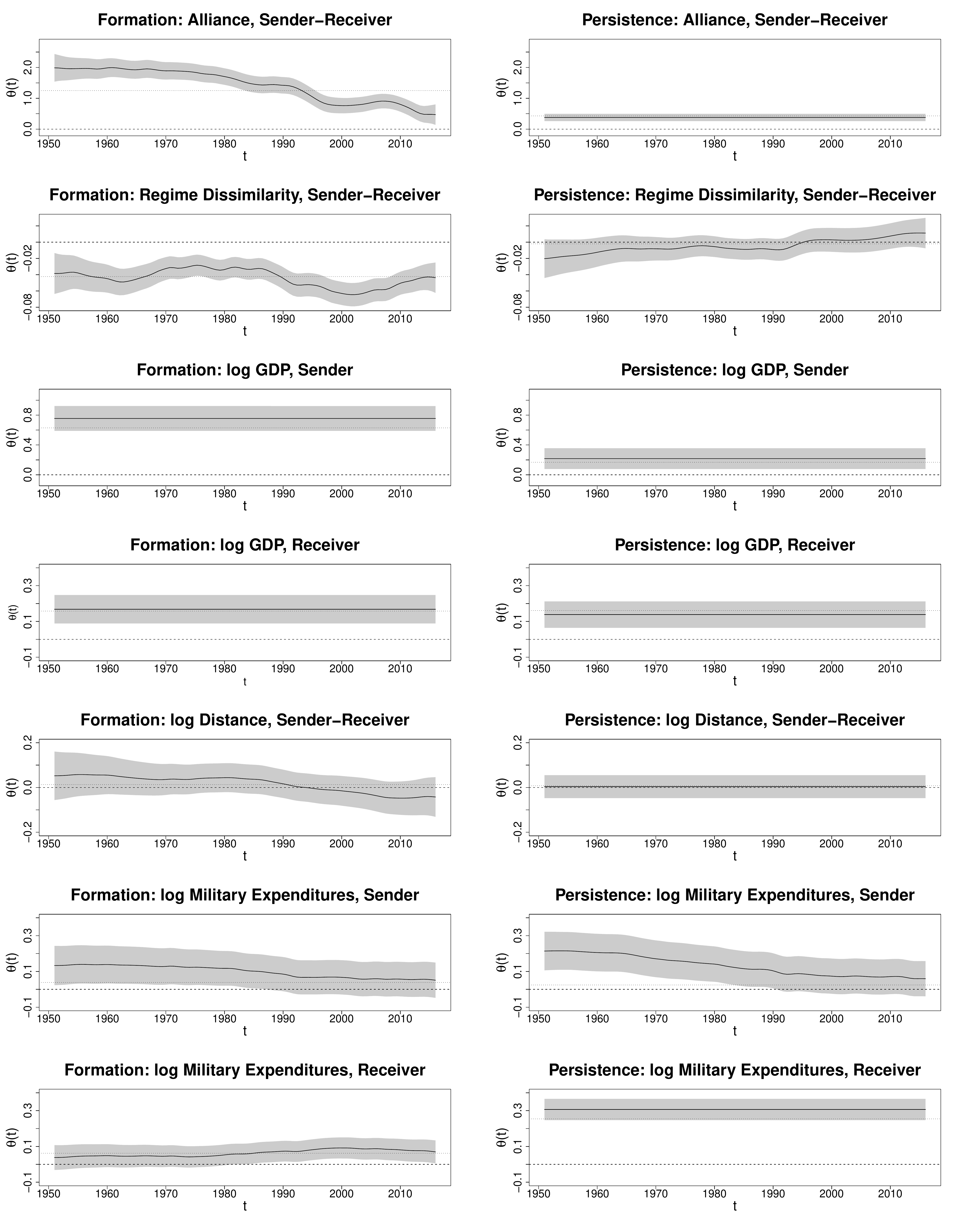}
	\caption{Time-varying coefficients of political and economic covariates in solid black. Shaded areas give two standard error bounds. Time-constant effects in dashed grey and zero line in dotted black.}
	\label{res_polecon}
\end{figure}

The results of the time-varying effects are grouped into network-related covariates (presented in Figure \ref{res_net}) and political and economic covariates (presented in Figure \ref{res_polecon}). The left columns give the coefficients for the formation model and the right columns for the persistence model, respectively. In the case of the network statistics,  a schematic representation of the corresponding network effects is added on the right hand side. The values for the coefficients are presented as solid lines with shaded regions, indicating two standard error bounds. The zero-line is indicated as dashed line and the estimates for time-constant coefficients are given by the dotted horizontal line. Note that the coefficients at a given time point can be interpreted just as the coefficients in a simple logit model. Additionally, for the same coefficient (or coefficients with the same norming) in the formation and persistence model, the effect size can be compared directly.

\vspace{0.2cm}	
\noindent{\sl Network-Effects} (see Figure \ref{res_net})

\noindent{\sl Outdegree}:
The senders' outdegree has a coefficient that is almost time-constant and close to zero for both models. This stands in contrast to the findings of \citet{Thurner2018}, where a strong effect is present. Hence, once controlled for country-specific heterogeneity (especially the sender-specific country effect), no population-level outdegree effect for the exporter is present (we show in the Supplementary Material that the effect is indeed present when country-specific heterogeneity is excluded).

However, the inclusion of country-specific sender and receiver effects does not affect the effect of the receivers' outdegree and the coefficient is consistently negative, and slightly increasing over time in the formation model. For the persistence model, we find a less pronounced but significant negative effect. We interpret this as clear evidence that countries with a high outdegree are comparatively less frequently importing, and importers usually have relatively less frequent export relations. According to our experience this specification captures the trade asymmetries of the oligopolistic market better than just specifying the indegrees of the receiver.   
\\
{\sl Reciprocity}:  Controlling for the distinguished asymmetrical nature of the weapons transfers, we identify a positive and significant impact of reciprocity in the formation model. Reciprocity in repeated transfers is only a relevant feature after the breakdown of the bipolar block structure. We conclude that the asymmetric structure is more present in persistent trade relations with importing countries that are typically dependent on big exporters.
\\
{\sl Transitivity}:
Looking at three-node statistics it can be seen that the variable transitivity has a positive impact on the formation and persistence. In the formation model, the effect is insignificant in the first years. This may be influenced by the clear hegemony of the United States and the Soviet Union, respectively, immediately after World War II which did not require a shared control over the recipient country, because the donor was powerful enough to secure the terms of a deal. In the 1980s middle power countries became technologically more advanced and especially in the West, they joined the US in delivering to other countries. The pronounced change between 1990 and 2010 can be explained by the break up of the two hostile blocs and the interruption of long-standing arm-trading partnerships leading to a fundamental reorganization until 2010 when the effect came back to the level of 1990. Although these arguments are also valid for the persistence model, we see that transitivity is 
less relevant for ongoing, repeated transfers (the time constant effect in the formation model has twice the size as the one in the persistence model). This impression is also strengthened by the fact that the coefficient is not subject to changes over time.
\\
{\sl Shared Suppliers}:
The coefficients related to the shared suppliers corroborate our expectation that many shared suppliers lead to the formation of transfers (positive and significant coefficient for the whole time period in the formation model). This indeed mirrors the phenomenon described above: there is a hierarchy of producing countries in the world. Receiver countries $i$ and $j$ should become acquainted with these technologies and should have similar levels of production capacities. This allows them to exchange arms.	Also, the act of receiving both from the same supplier means that this country places trust to both receivers - such that this facilitates trust giving one to another.
On the other hand, in the persistence model, the effect is indeed significantly negative and virtually zero from the 1975 on, showing that repetitive trading is not promoted by many shared suppliers.
\vspace{0.2cm}\\
\noindent	{\sl Covariate Effects} (see Figure \ref{res_polecon})

\noindent	{\sl Formal Alliance}: 
The impact of a bilateral formal alliances on the formation of a transfer is positive and significant for both, the formation and with a more modest effect for the persistence, corroborating our expectation that formal alliances are most relevant for the formation, i.e. by passing the required threshold of starting weapons transfers. The required  threshold of trustfulness to start seems to decline over time for the initiation. Hence, while formal alliances play a central role for arms trading after the second world war, the formation of arms trades is less and less influenced by the existence of a formal alliance by the sending and receiving state. However, given there exists an alliance, the impact (despite being smaller) continues to be relevant for repeated transfers. This is an important insight as we show for the first time that formalized alliance actually breed a dense web of arms transfers.	
\\
{\sl Regime Dissimilarity}: 	
For the formation model, the coefficient  on the absolute difference of the polity scores is all along negative, significant and shows some time variation.
With the decay of the eastern bloc, the resistance to send new arms to dissimilar regimes increases until 2000. After that, the absolute effect of different polity scores declines again, coming back to the long-term constant effect. Interestingly, we find that regime dissimilarity is irrelevant in the persistence model, showing that given a relationship is started, repetition does no more require regimes to exhibit shared governance values. 
\\
{\sl GDP}: 
As expected, the coefficients on the logarithmic GDP for sender and receiver are positive and constant for both models. However, the effect for the senders' GDP is much stronger in the formation model, showing that indeed mostly economically strong countries are able to open new markets for arms exports. Together, the coefficients  support the "gravity hypothesis", i.e.\ greater economic power and market sizes of the sender as well as the receiver increases the probability of forming and maintaining trade relations. However, given a transfer relation is started, this effect becomes smaller for repetition.
\\
{\sl Distance}:  In accordance with previous insights (\citealt{Thurner2018}), the results on the logarithmic distance contradicts the standard gravity model and distance proves to be insignificant in both models.
\\
{\sl Military Expenditures}:
For the military expenditures of the sender, we find very comparable and declining effects that become insignificant from 1990 on in both models. This indicates that with the end of the cold war the dominance of exporting countries with high military budgets has decreased. For the receivers' military expenditures in the formation model, the effect is positive and turns significant with time. This clearly illustrates that the military expenditures of the receiver are  not as important in the Cold War period where super powers often granted military assistance. Only with end of the 1980s there begins a marketization of the weapons transfers with suppliers demanding money for delivery. Given there is a preceding exchange, we find a very strong effect for the military expenditures of the receiver for the full observational period, indicating that  
the availability of huge military expenditures is a key for understanding the continuous yearly inflow of weapons. 

\vspace{0.2cm}
Overall, the results confirm our initial hypothesis. Judged by the size of the coefficients and their significance we find that the network statistics (reciprocity, transitivity, shared suppliers) and security related covariates (formal alliance, regime dissimilarity) prove to be highly influential in the formation model. On the other hand, we find weaker (or insignificant) network effects in the persistence model combined with a high dominance of the GDP and especially the military expenditures of the receiving country. This is not to say that we regard for example the positive effect of transitivity or alliances in the persistence model as irrelevant for repeated trading since the special nature of arms trading clearly demands trust for the formation and the persistence of transfers but the effects nevertheless show that the two processes are guided by different mechanisms that attach different priorities to security-related and economic variables. 

\FloatBarrier
\subsection{Time-varying smooth random effects}
\subsubsection{Functional component analysis}
\begin{figure}[t!]
	
	\centering
	\includegraphics[trim={0cm 0cm 0cm 0cm},clip,scale=0.188]{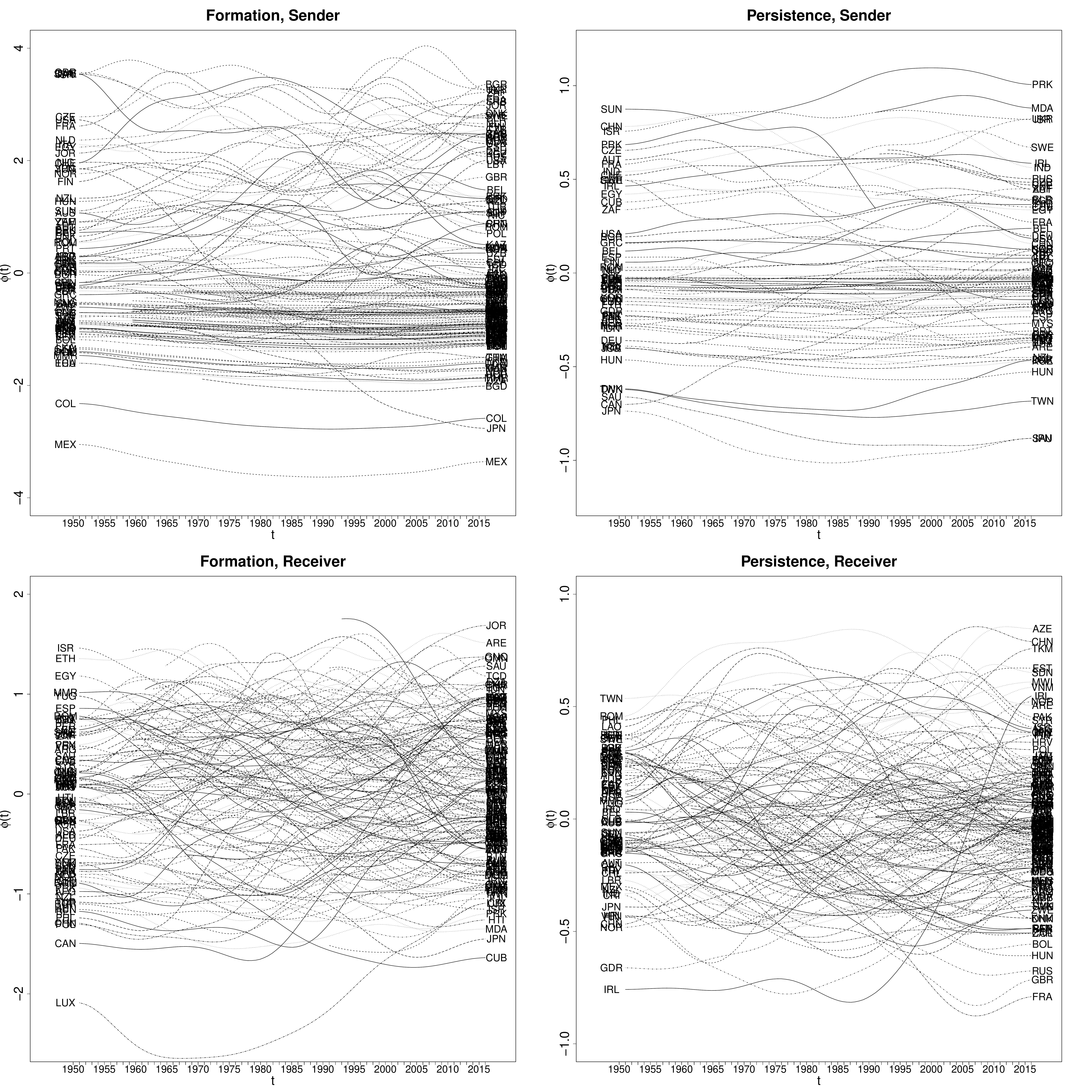}
	\caption{Fitted time-varying smooth random effects $\phi(t)$ plotted against time with country codes. The respective models are in the columns (formation on the left and persistence on the right) and the type of random effects in the rows (sender effect on the top and receiver effect on the bottom).}
	\label{smooth_1}
\end{figure}

We now pay attention to the actor-specific heterogeneity.  In Figure \ref{smooth_1}, the country-specific effects for the sender, as well as the receiver countries are visualized for the formation model on the left and the persistence model on the right. Note that in these plot we have truncated the curves for the years where countries are not existent.

\begin{figure}[t!]
	
	\centering
	\includegraphics[trim={0cm 0cm 0cm 0cm},clip,scale=0.28]{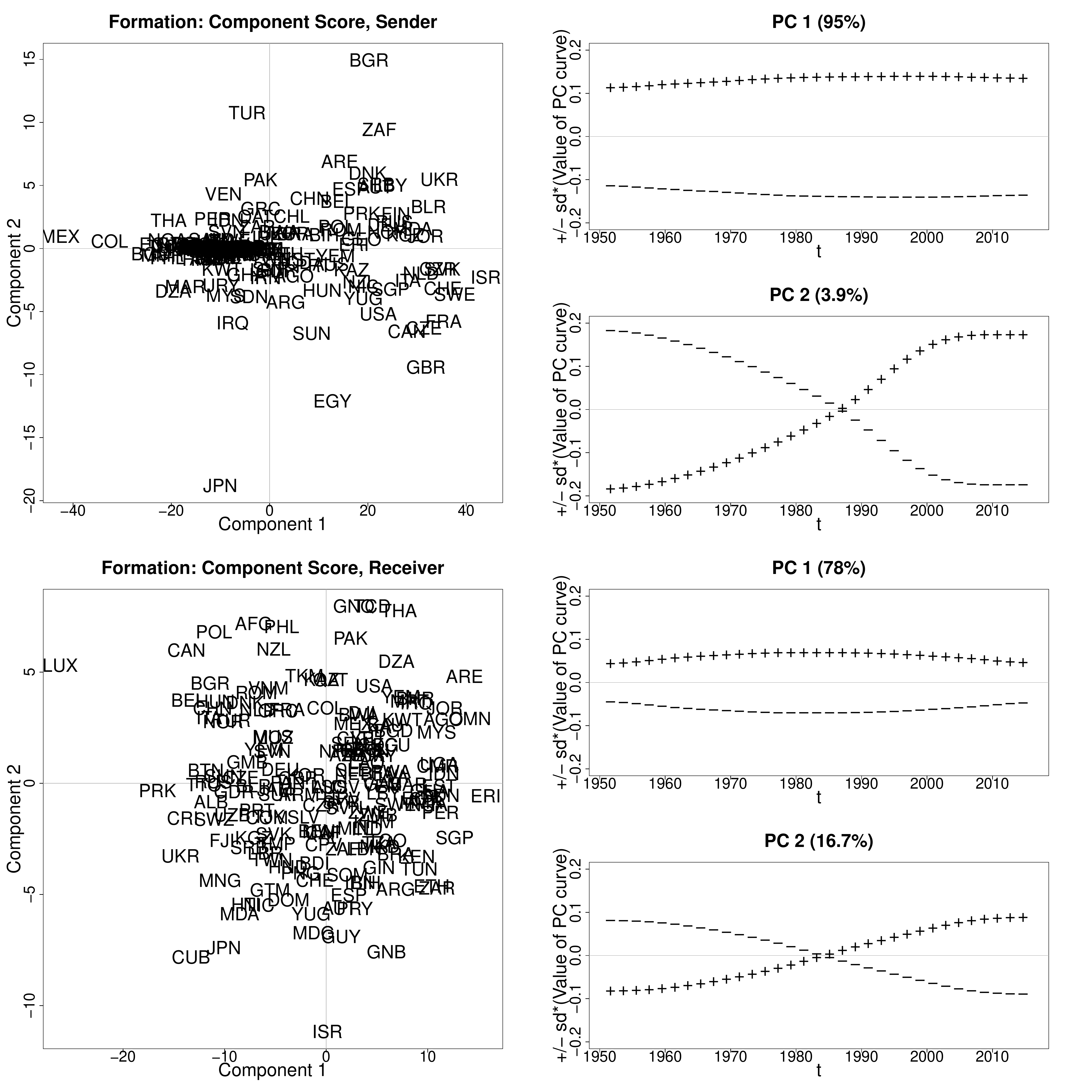}
	
	\caption{Functional principal component analysis of the smooth random effects in the formation model for the sender (top) and the receiver (bottom). Scores of the random effects for the first two principal components are given on the left. Mean principal component curve (zero line) and the effects of adding (+) and subtracting the principal component curve are given on the right.}
	\label{prin1}
\end{figure}

\begin{figure}[t!]
	\centering
	\includegraphics[trim={0cm 0cm 0cm 0cm},clip,scale=0.28]{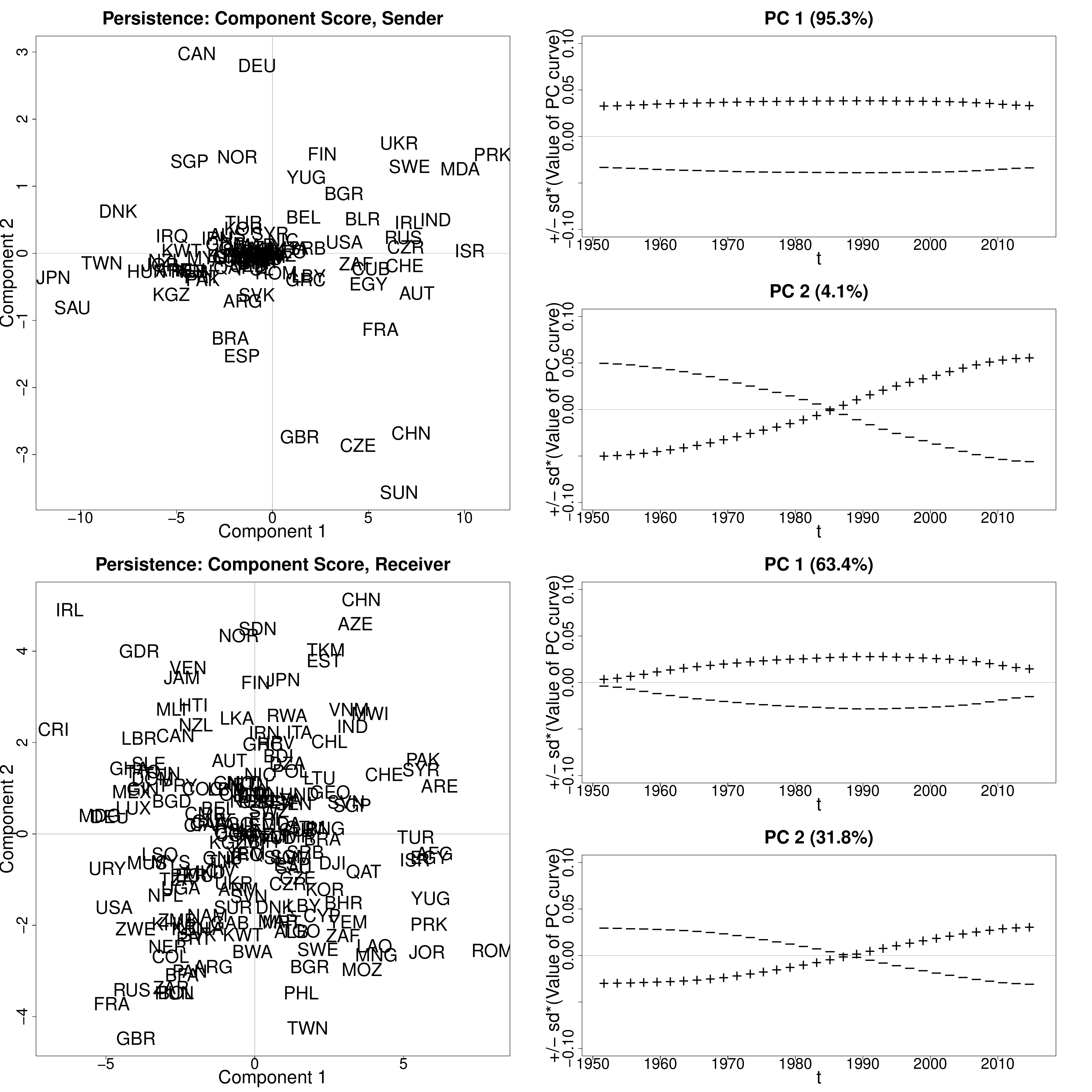}
	\caption{Functional principal component analysis of the smooth random effects in the persistence model for the sender (top) and the receiver (bottom). Scores of the random effects for the first two principal components are given on the left. Mean principal component curve (zero line) and the effects of adding (+) and subtracting the principal component curve are given on the right.}
	\label{prin2}
\end{figure}

At a first sight, interpretation of these plots looks clumsy. We therefore retrieve information by  employing a functional principal component analysis to the multivariate time series of random effects seen in Figure \ref{smooth_1} (see also \citealt{ramsay2005} and the Appendix \ref{detail_smooth}). The results are shown in Figure \ref{prin1} for the formation model and in Figure \ref{prin2} for the persistence model. On the left hand side the scores of the first two principal components are plotted, where the latter are visualized on the right hand side. The share of variance explained by the respective component is provided in the brackets. The basic idea of the approach is to show the effect of the principal components as perturbations from the mean random effects curves. By adding (the "+" line) or subtracting (the "-" line) a multiple of the principal component curve we get the visualized perturbation from the mean.

The first principal component is close to be constant and represents the share of variance induced by different overall levels of the random effect curves. The dynamic of the random effects is captured by the second principal component, delivering a tendency for an upward movement if positive and downward if negative. Hence, looking on the horizontal axes, we see countries that build up their arm trade links over the years as exporters (importers) on the right hand side while countries that are reluctant to building up export (import) links are plotted on the left hand side. Looking on the vertical axes, we see countries that decrease their role as exporter (importer) over the time on the bottom, and vice versa countries that increase the number of export (import) links over time on the top. All these effects are conditional on the remaining covariate effects discussed before. Hence, these random effects capture the remaining heterogeneity not included in the remaining model.

\subsubsection{Results of the functional component analysis}
Because of the great amount of information condensed in Figures \ref{prin1} and \ref{prin2} we restrict our interpretation to a few global patterns and selected countries that take either very special positions in the arms trade network (high or low values for component 1) or exhibit variation over time (high or low values for component 2).
Overall regarding the different levels of the random effects, it can already be seen in  Figure \ref{smooth_1}, that the heterogeneity is much more pronounced in the formation model in comparison to the persistence model. Furthermore, in the formation model, the countries differ more strongly in their ability to export in comparison to their ability to import while this contrast is not present in the persistence model.

A global pattern regarding the dynamics of the sender effect becomes visible since the top left in Figure \ref{prin1} looks like a lying mushroom. That is, countries that started on a low level (i.e.\ negative component 1) show, with the exception of Japan (JPN) and Turkey (TUR), not very much upward or downward variability (i.e.\ low level for component 2). In contrast, countries that have a random effect above zero move more strongly up or down with time. This means that the export dynamics are mainly driven by countries with relatively high sender effects.

Figures \ref{prin1} and \ref{prin2} show very well that fundamental changes of the system are driven by the end of the cold war. This can be seen exemplary regarding the position of the Soviet Union (SUN) and Czechoslovakia (CZE) in the top left in Figures \ref{prin1} and \ref{prin2} (both with a high level for component 1 and a low level for component 2). This mirrors that these countries left the system shortly after the collapse of the eastern bloc. However, this turning point affected not only exporters but also importers and consequently the representation of the receiver effects of the formation model at the bottom left of Figure \ref{prin1} is populated with (former) socialist countries such as Cuba (CUB), Ukraine (UKR), North Korea (PRK), Yugoslavia (YUG) and Moldova (MDA).
Additionally, we find a prominent position for Romania (ROM), being a country that has a high level (high value for component 1) but decreased its' tendency to be a receiver in persistent trade relations (low value for component 2) in Figure \ref{prin2}. However, while some of the countries of the eastern block ceased to exist or strongly reduced their exports or imports we also find a contrary pattern.    Countries like Ukraine (UKR) and Bulgaria (BGR) have managed to increase their sender effect in the formation as well as in the persistence model with time (high value for component 1 and component 2 in the top left of Figures \ref{prin1} and \ref{prin2}). This indicates that some left overs of the collapsed Soviet Union defence industries sold off their stocks and rushed into the global market of military products. 


Besides the massive shift initiated by the end of the cold war, we see that some dominant exporting countries, especially Great Britain (GBR), France (FRA) and Egypt (EGY), lost importance over time. This countries can be found in the fourth quadrant of the top left panels in  Figures \ref{prin1} and \ref{prin2}, meaning their  high sender effects decreased strongly with time. This might seem surprising since France and Great Britain are still among the countries with the highest exported volumes. However, France and Great Britain have left their dominance over former colonies  leading to a loss of control over many potential importers. The general pattern also carries over to their receiver effects.  Looking at the scores of Great Britain (GBR) and France (FRA) at the bottom left of Figure \ref{prin2} we see a strong decrease of their receiver effects in the persistence model.

Apart from global patterns, some countries exhibit exceptional scores that can be traced back to country-specific circumstances.
We find that Japan (JPN) stands out among the countries with the lowest proclivity to import (see the low scores for components 1 and 2 at the bottom left of Figure \ref{prin1}). Even more pronounced is the astonishing low tendency to export, mirrored by Japan's sender effect in the persistence models (Figure \ref{prin2}, top left) and the strongly declining sender effect in the formation model (Figure \ref{prin1}, top left). This stands in contrast to the fact that Japan is among the wealthiest countries with a highly developed export industry and is clearly due to the highly restrictive arms export principles introduced in 1967, and tightened in 1976. This ban on exports was only lifted in 2014 (\citealt{hughes2018}; \citealt{JAP2014}). 

Another, very notable case is Israel (ISR), being somehow the opposite pole in comparison to Japan (JPN). The sender effects on the top left of Figures \ref{prin1} and \ref{prin2} show that Israel (ISR) has an outstanding tendency to establish and maintain arms exports. On the other hand, Israel (ISR) takes a very polar position in the bottom left of Figure \ref{prin1} as a consequence of a strongly decreased (i.e.\ low level for component 2) receiver effect in the formation model.
These results reflect the country's path of developing highly internationally competitive weapons systems and its' rise to be one of the most important exporters. This stands in contrast to countries like Mexico (MEX), being the country with the least tendency to form new trade exports (top left in Figure \ref{prin1}). It appears that this country is not able to be a relevant player in the market despite being among the worlds' largest economies.  We consider these special paths as induced by cumulative advantages and learning over time in the one case (Israel), whereas in the case of Mexico (MEX) we observe the stickiness and path inertia of a country having not been able to make its defence products sold externally.

There remain many other interesting cases. For example the rise of  South Africa (ZAF) as an exporter in the formation model (top left in Figure \ref{prin1}), mirroring the history of the country, being initially dependent on imports and now among the major exporters of MCW. We also find that Ireland (IRL) strongly increased its tendency to be a persistent importer after its entry to the European Union (bottom left in Figure \ref{prin2}) while Germany (DEU) and Canada (CAN) strongly increased their roles as persistent exporters  (top left in Figure \ref{prin2}).

\FloatBarrier

\subsection{Model evaluation}

\begin{figure}[t]
	\centering

	\centering\includegraphics[trim={0cm 1cm 0cm 0cm},clip,width=0.95\textwidth]{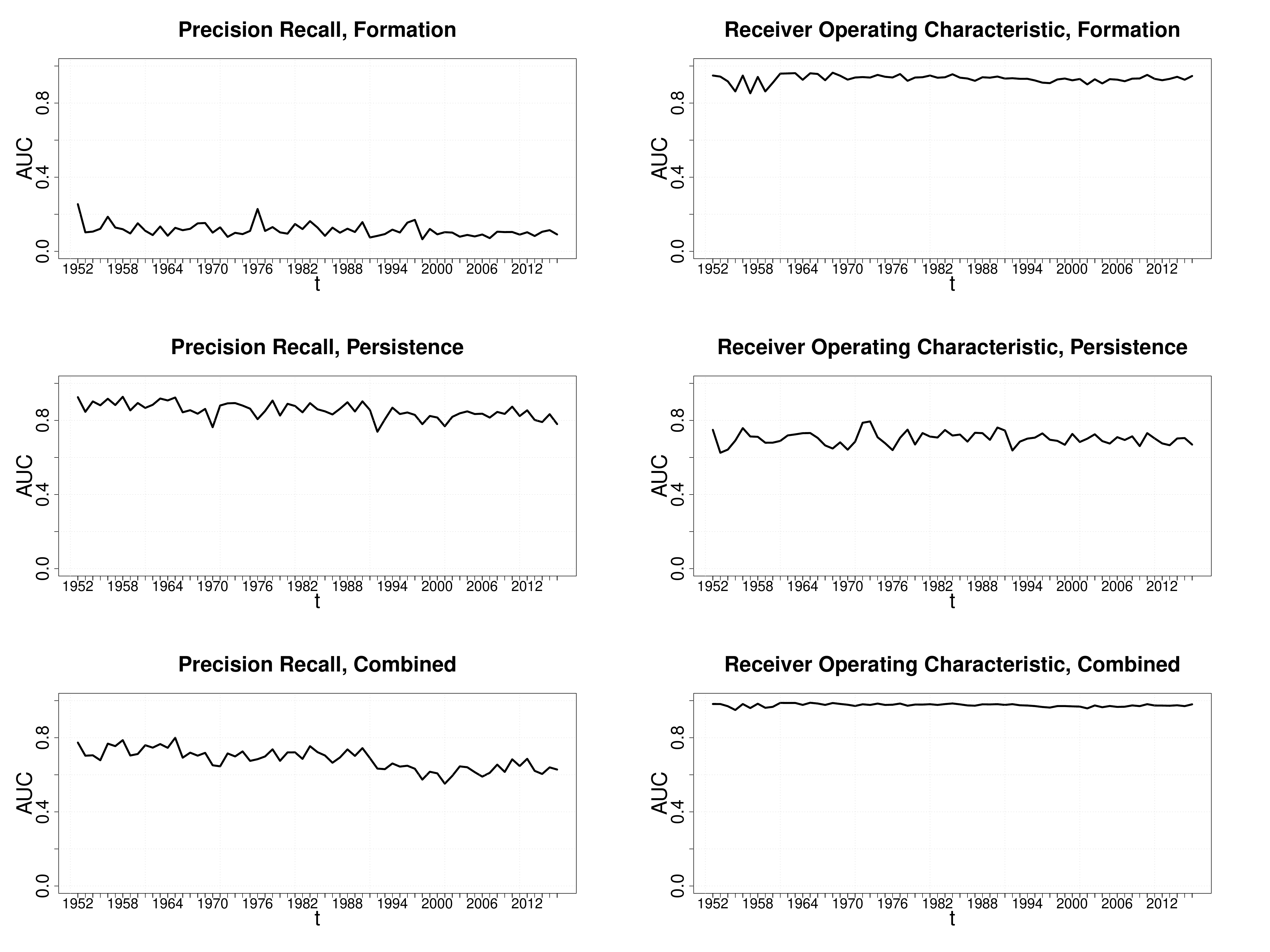}

	\caption{Time series of area under the curve (AUC) values for precision recall (PR) on the left and AUC values for the receiver operating characteristic (ROC) on the right. Formation model in the first row, persistence model in the second row and their combination in the last row.}
	\label{fig_combined}	
\end{figure}

The evaluation of the out-of-sample predictive power is based on the following steps. We first fit the formation model as well as the persistence model, based on the information in $t-1$, to the data in $t$ and use the estimated coefficients for the prediction of new formation or persistence of existing ties in $t+1$.	As the predictions are probabilistic by their nature, we weight the recall (true positive rate) against the false positive rate  for varying threshold levels, yielding the ROC curve and the AUC for each year of prediction. Because arms transfers can be regarded as rare events we also compute the PR curve and the corresponding AUC.
The results are plotted in Figure \ref{fig_combined} with the AUC values that correspond to the PR curves on the left and the one corresponding to the ROC curves on the right. The first row gives the evaluation of the formation model and the second row shows the persistence model. 
While the AUC values in the formation model are very high when evaluated at the ROC curves they are much lower with the PR curves. This is a consequence of being right quite frequently if a zero is predicted, while its is hard to forecast the actual transfers in the next period in case of the formation model. Interestingly, the opposite holds for the persistence model. In the combined version at the bottom of Figure \ref{fig_combined} the AUC values derived from the PR curve show that the model does quite well.

Additionally, we evaluate how well global network structures like the mean outdegree, the share of reciprocity and observed transitivity can be mirrored by the predictions using a simulation-based approach (see \citealt{Hunter2008a}). To do so, we fit the models for the transition between $t-1$ and $t$ and simulate from the formation model an the persistence model  $1,000$ times based on the information in $t$. Then, based on equation (\ref{construct}), the predicted network for $t+1$ is constructed. From this, we evaluate global network characteristics and compare them to the actual characteristics from the true MCW trade network in $t+1$. The corresponding Figure \ref{fig_sim1} is given in the Appendix \ref{osp_sim}. The results are reassuring and the simulated networks mirror the real network properties in an acceptable way. 

Clearly the proposed model is not the only suitable network model.  Alternatively, it is possible to analyse the data with a STERGM without random effects and with various variants of the ERGM or the TERGM with and without random effect. We discuss this extensively in the Supplementary Material and show that the out-of-sample predictive power of our model is superior to all other fitted candidate models. 

\section{Conclusion}\label{conc}
In this paper we employ a dynamic separable network model as introduced by \citet{krivitsky2014} and add techniques proposed by \citet{hastie1993} and \citet{durban2005}. This enables us to study the process of formation and persistence  separately as well as the inclusion of time-varying coefficients and smooth time-varying random effects that are further analysed by methods from functional data analysis as described in \citet{ramsay2005}.

Applied to the discretized MCW networks from 1950 to 2016 we find that the mechanisms leading to formation and persistence differ fundamentally. Most importantly, the formation is driven by network effects and security related variables, while the persistence of transfers is dominated the military expenditures of the receiving country. A careful analysis of the random effects exhibits a high variation among the countries as well as along the time dimension. By using functional principal component analysis we decompose the functional time series of smooth random effects in order find countries that have increased or decreased their relative importance in the network. The evaluation of the fit confirms that the chosen model is able to give good out-of-sample predictions.

	\newpage
\bibliographystyle{Chicago}
\bibliography{literature}

\begin{thebibliography}{}

\bibitem[\protect\citeauthoryear{Akerman and Seim}{Akerman and
  Seim}{2014}]{akerman2014}
Akerman, A. and A.~L. Seim (2014).
\newblock The global arms trade network 1950--2007.
\newblock {\em Journal of Comparative Economics\/}~{\em 42\/}(3), 535--551.

\bibitem[\protect\citeauthoryear{Almquist and Butts}{Almquist and
  Butts}{2014}]{almquist2014}
Almquist, Z.~W. and C.~T. Butts (2014).
\newblock Logistic network regression for scalable analysis of networks with
  joint edge/vertex dynamics.
\newblock {\em Sociological methodology\/}~{\em 44\/}(1), 273--321.

\bibitem[\protect\citeauthoryear{Barab{\'a}si and Albert}{Barab{\'a}si and
  Albert}{1999}]{barabasi1999}
Barab{\'a}si, A.-L. and R.~Albert (1999).
\newblock Emergence of scaling in random networks.
\newblock {\em Science\/}~{\em 286\/}(5439), 509--512.

\bibitem[\protect\citeauthoryear{Barigozzi, Fagiolo, and
  Garlaschelli}{Barigozzi et~al.}{2010}]{barigozzi2010}
Barigozzi, M., G.~Fagiolo, and D.~Garlaschelli (2010).
\newblock Multinetwork of international trade: A commodity-specific analysis.
\newblock {\em Physical Review E\/}~{\em 81\/}(4), 046104.

\bibitem[\protect\citeauthoryear{Blanton}{Blanton}{2005}]{Blanton2005}
Blanton, S.~L. (2005, 11).
\newblock {Foreign Policy in Transition? Human Rights, Democracy, and U.S. Arms
  Exports}.
\newblock {\em International Studies Quarterly\/}~{\em 49\/}(4), 647--667.

\bibitem[\protect\citeauthoryear{Block, Koskinen, Hollway, Steglich, and
  Stadtfeld}{Block et~al.}{2018}]{block2018}
Block, P., J.~Koskinen, J.~Hollway, C.~Steglich, and C.~Stadtfeld (2018).
\newblock Change we can believe in: Comparing longitudinal network models on
  consistency, interpretability and predictive power.
\newblock {\em Social Networks\/}~{\em 52}, 180 -- 191.

\bibitem[\protect\citeauthoryear{Bramoullé, Galeotti, Rogers, and
  Chaney}{Bramoullé et~al.}{2019}]{NetworksinInternationalTrade}
Bramoullé, Y., A.~Galeotti, B.~Rogers, and T.~Chaney (2019).
\newblock {\em Networks in International Trade}.
\newblock Oxford: Oxford University Press.

\bibitem[\protect\citeauthoryear{{Center for systemic Peace}}{{Center for
  systemic Peace}}{2017}]{Polity2016}
{Center for systemic Peace} (2017).
\newblock Polity {IV} annual time-series, 1800-2015, version 3.1.
\newblock Accessed: 2017-06-02.

\bibitem[\protect\citeauthoryear{{Correlates of War Project}}{{Correlates of
  War Project}}{2017a}]{Defagr2016}
{Correlates of War Project} (2017a).
\newblock International military alliances, 1648-2012, version 4.1.
\newblock Accessed: 2017-05-03.

\bibitem[\protect\citeauthoryear{{Correlates of War Project}}{{Correlates of
  War Project}}{2017b}]{cinc2017}
{Correlates of War Project} (2017b).
\newblock National material capabilities, 1816-2012, version 5.0.
\newblock Accessed: 2017-02-06.

\bibitem[\protect\citeauthoryear{Csardi and Nepusz}{Csardi and
  Nepusz}{2006}]{csardi2006}
Csardi, G. and T.~Nepusz (2006).
\newblock The igraph software package for complex network research.
\newblock {\em InterJournal, Complex Systems\/}~{\em 1695\/}(5), 1--9.

\bibitem[\protect\citeauthoryear{Disdier and Head}{Disdier and
  Head}{2008}]{disdier2008}
Disdier, A.-C. and K.~Head (2008).
\newblock The puzzling persistence of the distance effect on bilateral trade.
\newblock {\em The Review of Economics and Statistics\/}~{\em 90\/}(1), 37--48.

\bibitem[\protect\citeauthoryear{Duijn, Snijders, and Zijlstra}{Duijn
  et~al.}{2004}]{duijn2004}
Duijn, M.~A., T.~A. Snijders, and B.~J. Zijlstra (2004).
\newblock p2: a random effects model with covariates for directed graphs.
\newblock {\em Statistica Neerlandica\/}~{\em 58\/}(2), 234--254.

\bibitem[\protect\citeauthoryear{Durban and Aguilera-Morillo}{Durban and
  Aguilera-Morillo}{2017}]{durban2017}
Durban, M. and M.~C. Aguilera-Morillo (2017).
\newblock On the estimation of functional random effects.
\newblock {\em Statistical Modelling\/}~{\em 17\/}(1-2), 50--58.

\bibitem[\protect\citeauthoryear{Durb{\'a}n, Harezlak, Wand, and
  Carroll}{Durb{\'a}n et~al.}{2005}]{durban2005}
Durb{\'a}n, M., J.~Harezlak, M.~Wand, and R.~Carroll (2005).
\newblock Simple fitting of subject-specific curves for longitudinal data.
\newblock {\em Statistics in Medicine\/}~{\em 24\/}(8), 1153--1167.

\bibitem[\protect\citeauthoryear{Eilers and Marx}{Eilers and
  Marx}{1996}]{eilers1996}
Eilers, P.~H. and B.~D. Marx (1996).
\newblock Flexible smoothing with {B}-splines and penalties.
\newblock {\em Statistical Science\/}~{\em 11\/}(2), 89--102.

\bibitem[\protect\citeauthoryear{Erickson}{Erickson}{2015}]{eric2015}
Erickson, J.~L. (2015).
\newblock {\em Dangerous Trade: Arms Exports, Human Rights, and International
  Reputation}.
\newblock New York: Columbia University Press.

\bibitem[\protect\citeauthoryear{Garcia-Alonso and Levine}{Garcia-Alonso and
  Levine}{2007}]{garcia2007}
Garcia-Alonso, M.~D. and P.~Levine (2007).
\newblock Arms trade and arms races: A strategic analysis.
\newblock In T.~Sandler and K.~Hartley (Eds.), {\em Handbook of Defense
  Economics: Defense in a globalized world}, Volume~2, pp.\  941--971.
  Amsterdam: Elsevier Science Publishing.

\bibitem[\protect\citeauthoryear{Gleditsch}{Gleditsch}{2013a}]{gleditsch2013d}
Gleditsch, K.~S. (2013a).
\newblock Distance between capital cities.
\newblock Accessed: 2017-04-07.

\bibitem[\protect\citeauthoryear{Gleditsch}{Gleditsch}{2013b}]{gleditsch2013}
Gleditsch, K.~S. (2013b).
\newblock Expanded trade and {GDP} data.
\newblock Accessed: 2017-04-07.

\bibitem[\protect\citeauthoryear{Grau, Grosse, and Keilwagen}{Grau
  et~al.}{2015}]{grau2015}
Grau, J., I.~Grosse, and J.~Keilwagen (2015).
\newblock {PRROC}: computing and visualizing precision-recall and receiver
  operating characteristic curves in {R}.
\newblock {\em Bioinformatics\/}~{\em 31\/}(15), 2595--2597.

\bibitem[\protect\citeauthoryear{Handcock, Hunter, Butts, Goodreau, and
  Morris}{Handcock et~al.}{2008}]{handcock2008}
Handcock, M.~S., D.~R. Hunter, C.~T. Butts, S.~M. Goodreau, and M.~Morris
  (2008).
\newblock statnet: Software tools for the representation, visualization,
  analysis and simulation of network data.
\newblock {\em Journal of Statistical Software\/}~{\em 24\/}(1), 1548--7660.

\bibitem[\protect\citeauthoryear{Handcock, Raftery, and Tantrum}{Handcock
  et~al.}{2007}]{handcock2007}
Handcock, M.~S., A.~E. Raftery, and J.~M. Tantrum (2007).
\newblock Model-based clustering for social networks.
\newblock {\em J. R. Statist. Soc. A\/}~{\em 170\/}(2), 301--354.

\bibitem[\protect\citeauthoryear{Hanneke, Fu, Xing, et~al.}{Hanneke
  et~al.}{2010}]{hanneke2010}
Hanneke, S., W.~Fu, E.~P. Xing, et~al. (2010).
\newblock Discrete temporal models of social networks.
\newblock {\em Electronic Journal of Statistics\/}~{\em 4}, 585--605.

\bibitem[\protect\citeauthoryear{Harkavy}{Harkavy}{1975}]{Harkavy1975}
Harkavy, R.~E. (1975).
\newblock {\em The Arms Trade and International Systems.}
\newblock Cambridge: Cambridge University Press.

\bibitem[\protect\citeauthoryear{Hastie and Tibshirani}{Hastie and
  Tibshirani}{1987}]{hastie1990}
Hastie, T. and R.~Tibshirani (1987).
\newblock Generalized additive models: some applications.
\newblock {\em J. Am. Statist. Ass.\/}~{\em 82\/}(398), 371--386.

\bibitem[\protect\citeauthoryear{Hastie and Tibshirani}{Hastie and
  Tibshirani}{1993}]{hastie1993}
Hastie, T. and R.~Tibshirani (1993).
\newblock Varying-coefficient models.
\newblock {\em J. R. Statist. Soc. B\/}~{\em 55\/}(4), 757--796.

\bibitem[\protect\citeauthoryear{Head and Mayer}{Head and
  Mayer}{2014}]{head2013}
Head, K. and T.~Mayer (2014).
\newblock Gravity equations: Workhorse, toolkit, and cookbook.
\newblock In G.~Gopinath, E.~Helpman, and K.~Rogoff (Eds.), {\em Handbook of
  international economics}, Volume~4, pp.\  131--195. Amsterdam: Elsevier
  Science Publishing.

\bibitem[\protect\citeauthoryear{Hlavac}{Hlavac}{2013}]{hlavac2013}
Hlavac, M. (2013).
\newblock stargazer: Latex code and ascii text for well-formatted regression
  and summary statistics tables.

\bibitem[\protect\citeauthoryear{Hoff, Fosdick, Volfovsky, and He}{Hoff
  et~al.}{2015}]{amen2015}
Hoff, P., B.~Fosdick, A.~Volfovsky, and Y.~He (2015).
\newblock amen: Additive and multiplicative effects models for networks and
  relational data.
\newblock R package version 1.3.

\bibitem[\protect\citeauthoryear{Hoff, Raftery, and Handcock}{Hoff
  et~al.}{2002}]{hoff2002}
Hoff, P.~D., A.~E. Raftery, and M.~S. Handcock (2002).
\newblock Latent space approaches to social network analysis.
\newblock {\em J. Am. Statist. Ass.\/}~{\em 97\/}(460), 1090--1098.

\bibitem[\protect\citeauthoryear{Holland and Leinhardt}{Holland and
  Leinhardt}{1981}]{holland1981}
Holland, P.~W. and S.~Leinhardt (1981).
\newblock An exponential family of probability distributions for directed
  graphs.
\newblock {\em J. Am. Statist. Ass.\/}~{\em 76\/}(373), 33--50.

\bibitem[\protect\citeauthoryear{Holme}{Holme}{2015}]{holme2015}
Holme, P. (2015).
\newblock Modern temporal network theory: a colloquium.
\newblock {\em The European Physical Journal B\/}~{\em 88\/}(9), 1--30.

\bibitem[\protect\citeauthoryear{Hughes}{Hughes}{2018}]{hughes2018}
Hughes, C. (2018).
\newblock Japan's emerging arms transfer strategy: diversifying to re-centre on
  the us–japan alliance.
\newblock {\em The Pacific Review\/}~{\em 31\/}(4), 424--440.

\bibitem[\protect\citeauthoryear{Hunter, Goodreau, and Handcock}{Hunter
  et~al.}{2008}]{Hunter2008a}
Hunter, D.~R., S.~M. Goodreau, and M.~S. Handcock (2008).
\newblock Goodness of fit of social network models.
\newblock {\em J. Am. Statist. Ass.\/}~{\em 103\/}(481), 248--258.

\bibitem[\protect\citeauthoryear{Koskinen, Caimo, and Lomi}{Koskinen
  et~al.}{2015}]{koskinen2015}
Koskinen, J., A.~Caimo, and A.~Lomi (2015).
\newblock Simultaneous modeling of initial conditions and time heterogeneity in
  dynamic networks: An application to foreign direct investments.
\newblock {\em Network Science\/}~{\em 3\/}(1), 58--77.

\bibitem[\protect\citeauthoryear{Krause}{Krause}{1995}]{krause1995arms}
Krause, K. (1995).
\newblock {\em Arms and the state: patterns of military production and trade}.
\newblock Cambridge: Cambridge University Press.

\bibitem[\protect\citeauthoryear{Krivitsky and Handcock}{Krivitsky and
  Handcock}{2014}]{krivitsky2014}
Krivitsky, P.~N. and M.~S. Handcock (2014).
\newblock A separable model for dynamic networks.
\newblock {\em J. R. Statist. Soc. B\/}~{\em 76\/}(1), 29--46.

\bibitem[\protect\citeauthoryear{Leifeld, Cranmer, and Desmarais}{Leifeld
  et~al.}{2018}]{leifeld2017}
Leifeld, P., S.~J. Cranmer, and B.~A. Desmarais (2018).
\newblock Temporal exponential random graph models with btergm: estimation and
  bootstrap confidence intervals.
\newblock {\em Journal of Statistical Software\/}~{\em 83\/}(6).

\bibitem[\protect\citeauthoryear{Lusher, Koskinen, and Robins}{Lusher
  et~al.}{2012}]{lusher2012}
Lusher, D., J.~Koskinen, and G.~Robins (2012).
\newblock {\em Exponential random graph models for social networks: Theory,
  methods, and applications}.
\newblock Cambridge: Cambridge University Press.

\bibitem[\protect\citeauthoryear{Marshall}{Marshall}{2017}]{marshall2017}
Marshall, M.~G. (2017).
\newblock Polity {IV} project: Political regime characteristics and
  transitions, 1800-2016.
\newblock Accessed: 2017-06-02.

\bibitem[\protect\citeauthoryear{{Ministry of Foreign Affairs of
  Japan}}{{Ministry of Foreign Affairs of Japan}}{2014}]{JAP2014}
{Ministry of Foreign Affairs of Japan} (2014).
\newblock Japan's policies on the control of arms exports.
\newblock Accessed: 2017-02-21.

\bibitem[\protect\citeauthoryear{Moritz}{Moritz}{2016}]{Moritz2016}
Moritz, S. (2016).
\newblock {\em imputeTS: Time series missing value imputation}.
\newblock R package version 2.6.

\bibitem[\protect\citeauthoryear{{R Development Core Team}}{{R Development Core
  Team}}{2008}]{team2014r}
{R Development Core Team} (2008).
\newblock {\em R: A Language and Environment for Statistical Computing}.
\newblock Vienna, Austria: R Foundation for Statistical Computing.

\bibitem[\protect\citeauthoryear{Ramsay and Silverman}{Ramsay and
  Silverman}{2005}]{ramsay2005}
Ramsay, J.~O. and B.~W. Silverman (2005).
\newblock {\em Functional data analysis}.
\newblock New York: Springer Science \& Business Media.

\bibitem[\protect\citeauthoryear{Robins and Pattison}{Robins and
  Pattison}{2001}]{robins2001}
Robins, G. and P.~Pattison (2001).
\newblock Random graph models for temporal processes in social networks.
\newblock {\em Journal of Mathematical Sociology\/}~{\em 25\/}(1), 5--41.

\bibitem[\protect\citeauthoryear{Ruppert, Wand, and Carroll}{Ruppert
  et~al.}{2009}]{ruppert2009}
Ruppert, D., M.~Wand, and R.~J. Carroll (2009).
\newblock Semiparametric regression during 2003--2007.
\newblock {\em Electronic Journal of Statistics\/}~{\em 1\/}(3), 1193--1256.

\bibitem[\protect\citeauthoryear{Schulze, Pamp, and Thurner}{Schulze
  et~al.}{2017}]{Schulze2017}
Schulze, C., O.~Pamp, and P.~W. Thurner (2017, 10).
\newblock {Economic Incentives and the Effectiveness of Nonproliferation Norms:
  German Major Conventional Arms Transfers 1953–2013}.
\newblock {\em International Studies Quarterly\/}~{\em 61\/}(3), 529--543.

\bibitem[\protect\citeauthoryear{Schweitzer, Fagiolo, Sornette, Vega-Redondo,
  Vespignani, and White}{Schweitzer et~al.}{2009}]{schweitzer2009}
Schweitzer, F., G.~Fagiolo, D.~Sornette, F.~Vega-Redondo, A.~Vespignani, and
  D.~R. White (2009).
\newblock Economic networks: The new challenges.
\newblock {\em Science\/}~{\em 325\/}(5939), 422--425.

\bibitem[\protect\citeauthoryear{Singer, Bremer, and Stuckey}{Singer
  et~al.}{1972}]{singer1972}
Singer, J.~D., S.~Bremer, and J.~Stuckey (1972).
\newblock Capability distribution, uncertainty, and major power war, 1820-1965.
\newblock {\em Peace, War, and Numbers\/}~{\em 19}, 19--48.

\bibitem[\protect\citeauthoryear{{SIPRI}}{{SIPRI}}{2017a}]{sipridata2017}
{SIPRI} (2017a).
\newblock Arms transfers database.
\newblock Accessed: 2017-06-02.

\bibitem[\protect\citeauthoryear{{SIPRI}}{{SIPRI}}{2017b}]{siprimeth2017}
{SIPRI} (2017b).
\newblock Arms transfers database - methodology.
\newblock Accessed: 2017-03-23.

\bibitem[\protect\citeauthoryear{Snijders}{Snijders}{2011}]{snijders2011}
Snijders, T.~A. (2011).
\newblock Statistical models for social networks.
\newblock {\em Annual Review of Sociology\/}~{\em 37\/}(1), 131--153.

\bibitem[\protect\citeauthoryear{Snijders, Koskinen, and
  Schweinberger}{Snijders et~al.}{2010}]{snijders2010}
Snijders, T.~A., J.~Koskinen, and M.~Schweinberger (2010).
\newblock Maximum likelihood estimation for social network dynamics.
\newblock {\em The Annals of Applied Statistics\/}~{\em 4\/}(2), 567.

\bibitem[\protect\citeauthoryear{Snijders, Van~de Bunt, and Steglich}{Snijders
  et~al.}{2010}]{snijders2010soam}
Snijders, T.~A., G.~G. Van~de Bunt, and C.~E. Steglich (2010).
\newblock Introduction to stochastic actor-based models for network dynamics.
\newblock {\em Social Networks\/}~{\em 32\/}(1), 44--60.

\bibitem[\protect\citeauthoryear{Squartini, Fagiolo, and
  Garlaschelli}{Squartini et~al.}{2011a}]{squartini2011a}
Squartini, T., G.~Fagiolo, and D.~Garlaschelli (2011a).
\newblock Randomizing world trade. {I. A} binary network analysis.
\newblock {\em Physical Review E\/}~{\em 84\/}(4), 046117.

\bibitem[\protect\citeauthoryear{Squartini, Fagiolo, and
  Garlaschelli}{Squartini et~al.}{2011b}]{squartini2011b}
Squartini, T., G.~Fagiolo, and D.~Garlaschelli (2011b).
\newblock Randomizing world trade. {II. A} weighted network analysis.
\newblock {\em Physical Review E\/}~{\em 84\/}(4), 046118.

\bibitem[\protect\citeauthoryear{Thiemichen, Friel, Caimo, and
  Kauermann}{Thiemichen et~al.}{2016}]{thiemichen2016}
Thiemichen, S., N.~Friel, A.~Caimo, and G.~Kauermann (2016).
\newblock Bayesian exponential random graph models with nodal random effects.
\newblock {\em Social Networks\/}~{\em 46}, 11--28.

\bibitem[\protect\citeauthoryear{Thurner, Schmid, Christian, and
  Kauermann}{Thurner et~al.}{2018}]{Thurner2018}
Thurner, P.~W., Schmid, C.~Christian, Skyler, and G.~Kauermann (2018).
\newblock Network interdependencies and the evolution of the international arms
  trade.
\newblock Online First: Journal of Conflict Resolution.

\bibitem[\protect\citeauthoryear{Wood}{Wood}{2006}]{wood2006}
Wood, S.~N. (2006).
\newblock Low-rank scale-invariant tensor product smooths for generalized
  additive mixed models.
\newblock {\em Biometrics\/}~{\em 62\/}(4), 1025--1036.

\bibitem[\protect\citeauthoryear{Wood}{Wood}{2011}]{Wood2011}
Wood, S.~N. (2011).
\newblock Fast stable restricted maximum likelihood and marginal likelihood
  estimation of semiparametric generalized linear models.
\newblock {\em J. R. Statist. Soc. B\/}~{\em 73\/}(1), 3--36.

\bibitem[\protect\citeauthoryear{Wood}{Wood}{2017}]{wood2017}
Wood, S.~N. (2017).
\newblock {\em Generalized additive models: an introduction with R}.
\newblock Boca Raton: CRC press.

\bibitem[\protect\citeauthoryear{Wood, Goude, and Shaw}{Wood
  et~al.}{2015}]{wood2015}
Wood, S.~N., Y.~Goude, and S.~Shaw (2015).
\newblock Generalized additive models for large data sets.
\newblock {\em J. R. Statist. Soc. C\/}~{\em 64\/}(1), 139--155.

\bibitem[\protect\citeauthoryear{{World Bank}}{{World Bank}}{2017}]{GDP2017}
{World Bank} (2017).
\newblock World bank open data, real {GDP}.
\newblock Accessed: 2017-04-01.

\bibitem[\protect\citeauthoryear{Xu}{Xu}{2015}]{xu2015}
Xu, K. (2015).
\newblock Stochastic block transition models for dynamic networks.
\newblock In {\em Proceedings of the 18th International Conference on
  Artificial Intelligence and Statistics (AISTATS)}, Volume~18, pp.\
  1079--1087.

\end{thebibliography}
\newpage
\newpage
\pagenumbering{Roman}
\appendix
\section{Appendix}

\subsection{Descriptives}\label{descrannex}
\begin{figure}[t!]
	\centering	
	\includegraphics[trim={5cm 4cm 5cm 4cm},clip,scale=0.9]{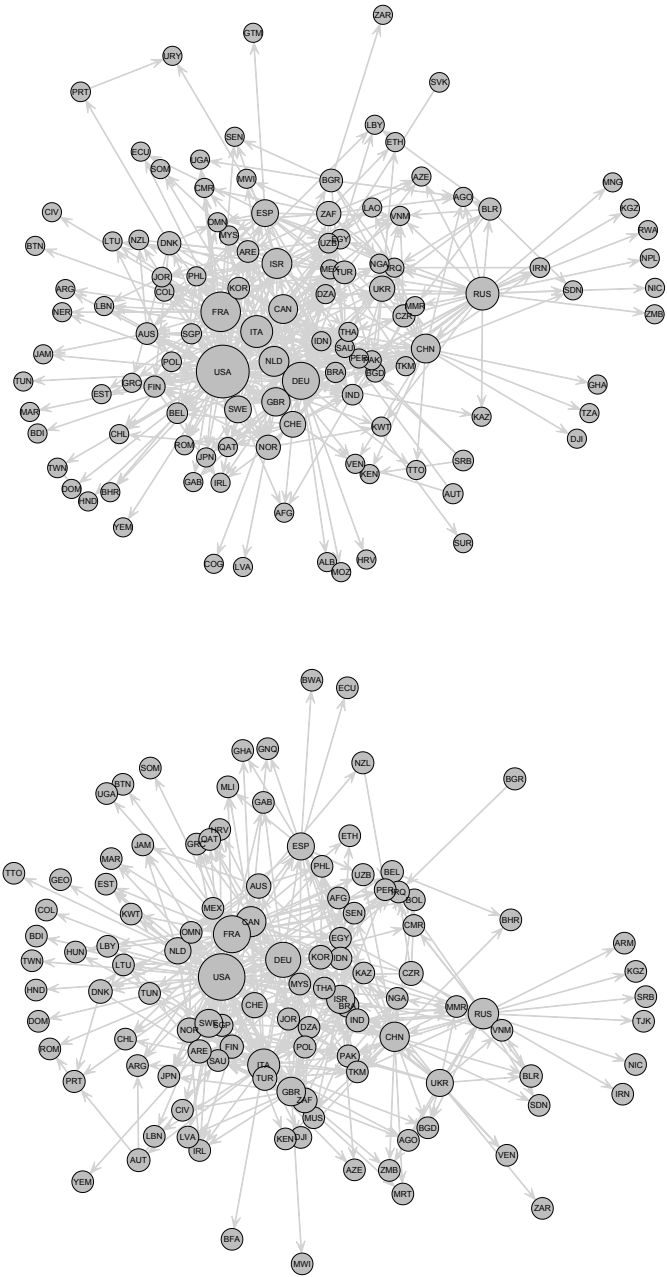}
	\caption{Network of international transfers of major conventional weapons (MCW) in 2015 (top) and 2016 (bottom). Countries are represented by vertices  and directed edges represent arms exports. Vertex sizes are scaled proportional to the logarithmic outdegree (number of outgoing edges).}
	\label{Fig-network-1}
\end{figure}
In Figure \ref{Fig-network-1}, the binary network is shown for the years 2015 and 2016. Table \ref{coverage} gives the categories of arms that are included in the analysis. All types with explanations are taken from \citet{siprimeth2017}. The 171 countries that are included in our analysis can be found in Table \ref{count_inc}, together with the three-digit country codes that are used to abbreviate countries in the paper. In  addition to that, the time periods, for which we coded the countries as existent are included. Note that the SIPRI data set contains more than 171 arm trading entities but we excluded non-states and countries with no (reliable) covariates available. In the covariate GDP some missings are present in the data. No time series of covariates for the selected countries is completely missing (those countries are excluded from the analysis) and the major share of them is complete but there are series with some missing values. This is  sometimes the case in the year $1990$ and/or $1991$ where the former socialist countries splitted up or had some transition time. In other cases values at the beginning or at the end of the series are missing. We have decided on three general rules to fill the gaps: First, if a value for a certain country is missing in $t$ but there are values available in $t-1$ and $t+1$, the mean of those is used. If the values are missing at the end of the observational period, the last value observed is taken. In case of missing values in the beginning, the first value observed is taken.

The series on military expenditures are imputed similarly, using linear interpolation by employing the \texttt{R} package \texttt{imputeTS} by \citet{Moritz2016}.

\begin{figure}[t]
	
	\centering
	\includegraphics[trim={0cm 0cm 0cm 0cm},clip,scale=0.50]{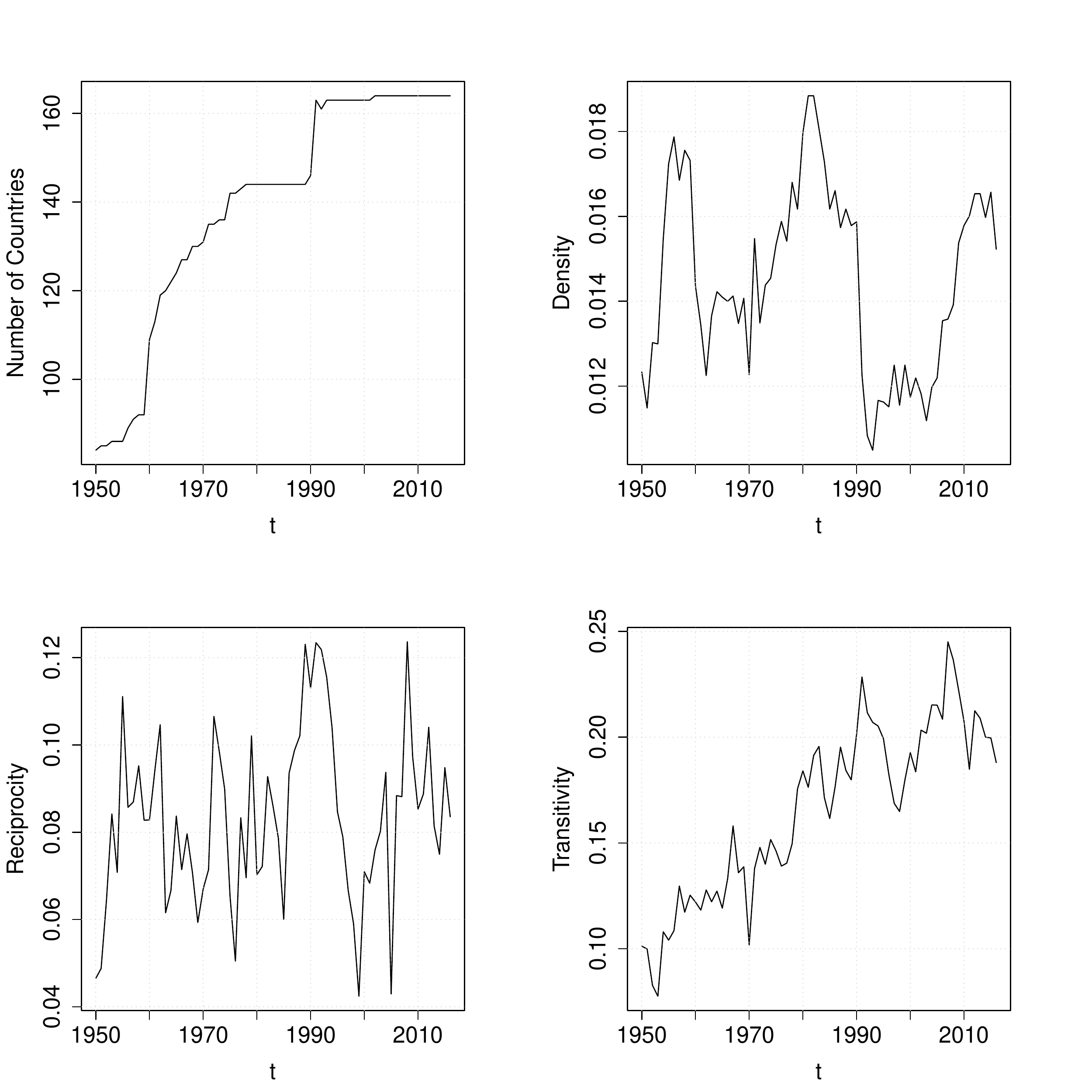}
	\caption{Time series from 1950 to 2016 of global network statistics for the international arms trade network for the included countries. Number of countries included on the top left panel. Number of realized transfers relative to the number of possible transfers on the top right Panel. Share of reciprocated transfers on the bottom left panel. Ratio of undirected triangles relative to connected triples on the bottom right.}
	\label{sum_stat}
\end{figure}

The number of countries included each year in the network is provided in the upper left panel of Figure \ref{sum_stat}. It can be seen that the network is growing almost each year until 1992, with two big leaps that show the effects of the decolonization, beginning in 1960 and the end of the Soviet Union after 1991. Typical descriptive statistics for the analysis of networks are {\sl Density}, {\sl Reciprocity} and {\sl Transitivity}, all shown in Figure \ref{sum_stat}. The {\sl Density} is defined as the  number of edges divided by the number of possible edges. {\sl Reciprocity} is defined as the share of trade flows being reciprocal.  {\sl Transitivity} is defined as the ratio of triangles and connected triples in the graph.

\FloatBarrier
\begin{table}[t!]
	\centering \scriptsize

	\begin{tabular}{l p{11cm}}
		
		\hline
		Type & Explanation \\ 
		\hline
		Aircraft                                                                                                                                                   & All fixed-wing aircraft and helicopters, including unmanned aircraft  with a minimum loaded weight of 20 kg. Exceptions are microlight aircraft, powered and unpowered gliders and target drones.                                                                                                                                                                                                                                                                                                       \\ 
		Air-defence systems                                                                                                                                         & (a) All land-based surface-to-air missile systems, and (b) all anti-aircraft guns with a calibre of more than 40 mm or with multiple barrels with a combined caliber of at least 70 mm. This includes self-propelled systems on armoured or unarmoured chassis.                                                                                                                                                                                                                                            \\
		Anti-submarine warfare weapons                                                                                                                             & Rocket launchers, multiple rocket launchers and mortars for use against submarines, with a calibre equal to or above 100 mm.                                                                                                                                                                                                                                                                                                                                                                                      \\
		Armoured vehicles                                                                                                                                          & All vehicles with integral armour protection, including all types of tank, tank destroyer, armoured car, armoured personnel carrier, armoured support vehicle and infantry fighting vehicle. Vehicles with very light armour protection (such as trucks with an integral but lightly armoured cabin) are excluded.                                                                                                                                                                                               \\
		Artillery                                                                                                                                                  & Naval, fixed, self-propelled and towed guns, howitzers, multiple rocket launchers and mortars, with a calibre equal to or above 100 mm.                                                                                                                                                                                                                                                                                                                                                                        \\
		Engines                                                                                                                                                    & (a) Engines for military aircraft, for example, combat-capable aircraft, larger military transport and support aircraft, including large helicopters; (b) Engines for combat ships -,fast attack craft, corvettes, frigates, destroyers, cruisers, aircraft carriers and submarines; (c) Engines for most armoured vehicles - generally engines of more than 200 horsepower output.$^{*}$                                                                                                                            \\
		Missiles                                                                                                                                                   & (a) All powered, guided missiles and torpedoes, and (b) all unpowered but guided bombs and shells. This includes man-portable air defence systems  and portable guided anti-tank missiles. Unguided rockets, free-fall aerial munitions, anti-submarine rockets and target drones are excluded.                                                                                                                                                                                                         \\
		Sensors                                                                                                                                                    & (a) All land-, aircraft- and ship-based active (radar) and passive (e.g.\ electro-optical) surveillance systems with a range of at least 25 kilometres, with the exception of navigation and weather radars, (b) all fire-control radars, with the exception of range-only radars, and (c) anti-submarine warfare and anti-ship sonar systems for ships and helicopters.$^{*}$                                                                                                                                                                                                                                                                                                                                                                                                                                                                                                                                                                                 \\
		Ships                                                                                                                                                      & (a) All ships with a standard tonnage of 100 tonnes or more, and (b) all ships armed with artillery of 100-mm calibre or more, torpedoes or guided missiles, and (c) all ships below 100 tonnes where the maximum speed (in kmh) multiplied with the full tonnage equals 3500 or more. Exceptions are most survey ships, tugs and some transport ships                                                                                                                                                           \\
		Other                                                                                                                                                      & (a) All turrets for armoured vehicles fitted with a gun of at least 12.7 mm calibre or with guided anti-tank missiles, (b) all turrets for ships fitted with a gun of at least 57-mm calibre, and (c) all turrets for ships fitted with multiple guns with a combined calibre of at least 57 mm, and (d) air refueling systems as used on tanker aircraft.$^{*}$                                                                                                                                                     \\ \hline
		\multicolumn{2}{p{14.2cm}}{$^{*}$In cases where the system is fitted on a platform (vehicle, aircraft or ship), the database only includes those systems that come from a different supplier from the supplier of the platform.}\\
		\multicolumn{2}{p{14.2cm}}{The Arms Transfers Database does not cover other military equipment such as small arms and light weapons (SALW) other than portable guided missiles such as man-portable air defence systems and guided anti-tank missiles. Trucks, artillery under 100-mm calibre, ammunition, support equipment and components (other than those mentioned above), repair and support services or technology transfers are also not included in the database.}
		\\
		\multicolumn{2}{p{16cm}}{Source: \citet{siprimeth2017}}
	\end{tabular}
	\caption{Types of weapon systems included in the SIPRI arms trade database.}
	\label{coverage}
\end{table}
\FloatBarrier

\begin{table}[ht]
	\centering \tiny
	\begin{tabular}{lllllllll}
		\hline
		Country & Code & Included & Country & Code & Included &Country & Code & Included \\  
		\hline
		Afghanistan & AFG & 1950 - 2016 & German Dem. Rep. & GDR & 1950 - 1991 & Pakistan & PAK & 1950 - 2016 \\ 
		Albania & ALB & 1950 - 2016 & Germany & DEU & 1950 - 2016 & Panama & PAN & 1950 - 2016 \\ 
		Algeria & DZA & 1962 - 2016 & Ghana & GHA & 1957 - 2016 & Papua New Guin. & PNG & 1975 - 2016 \\ 
		Angola & AGO & 1975 - 2016 & Greece & GRC & 1950 - 2016 & Paraguay & PRY & 1950 - 2016 \\ 
		Argentina & ARG & 1950 - 2016 & Guatemala & GTM & 1950 - 2016 & Peru & PER & 1950 - 2016 \\ 
		Armenia & ARM & 1991 - 2016 & Guinea & GIN & 1958 - 2016 & Philippines & PHL & 1950 - 2016 \\ 
		Australia & AUS & 1950 - 2016 & Guinea-Bissau & GNB & 1973 - 2016 & Poland & POL & 1950 - 2016 \\ 
		Austria & AUT & 1950 - 2016 & Guyana & GUY & 1966 - 2016 & Portugal & PRT & 1950 - 2016 \\ 
		Azerbaijan & AZE & 1991 - 2016 & Haiti & HTI & 1950 - 2016 & Qatar & QAT & 1971 - 2016 \\ 
		Bahrain & BHR & 1971 - 2016 & Honduras & HND & 1950 - 2016 & Romania & ROM & 1950 - 2016 \\ 
		Bangladesh & BGD & 1971 - 2016 & Hungary & HUN & 1950 - 2016 & Russia & RUS & 1992 - 2016 \\ 
		Belarus & BLR & 1991 - 2016 & India & IND & 1950 - 2016 & Rwanda & RWA & 1962 - 2016 \\ 
		Belgium & BEL & 1950 - 2016 & Indonesia & IDN & 1950 - 2016 & Saudi Arabia & SAU & 1950 - 2016 \\ 
		Benin & BEN & 1961 - 2016 & Iran & IRN & 1950 - 2016 & Senegal & SEN & 1960 - 2016 \\ 
		Bhutan & BTN & 1950 - 2016 & Iraq & IRQ & 1950 - 2016 & Serbia & SRB & 1992 - 2016 \\ 
		Bolivia & BOL & 1950 - 2016 & Ireland & IRL & 1950 - 2016 & Sierra Leone & SLE & 1961 - 2016 \\ 
		Bosnia Herzegov. & BIH & 1992 - 2016 & Israel & ISR & 1950 - 2016 & Singapore & SGP & 1965 - 2016 \\ 
		Botswana & BWA & 1966 - 2016 & Italy & ITA & 1950 - 2016 & Slovakia & SVK & 1993 - 2016 \\ 
		Brazil & BRA & 1950 - 2016 & Jamaica & JAM & 1962 - 2016 & Slovenia & SVN & 1991 - 2016 \\ 
		Bulgaria & BGR & 1950 - 2016 & Japan & JPN & 1950 - 2016 & Solomon Islands & SLB & 1978 - 2016 \\ 
		Burkina Faso & BFA & 1960 - 2016 & Jordan & JOR & 1950 - 2016 & Somalia & SOM & 1960 - 2016 \\ 
		Burundi & BDI & 1962 - 2016 & Kazakhstan & KAZ & 1991 - 2016 & South Africa & ZAF & 1950 - 2016 \\ 
		Cambodia & KHM & 1953 - 2016 & Kenya & KEN & 1963 - 2016 & Soviet Union & SUN & 1950 - 1991 \\ 
		Cameroon & CMR & 1960 - 2016 & North Korea & PRK & 1950 - 2016 & Spain & ESP & 1950 - 2016 \\ 
		Canada & CAN & 1950 - 2016 & South Korea & KOR & 1950 - 2016 & Sri Lanka & LKA & 1950 - 2016 \\ 
		Cape Verde & CPV & 1975 - 2016 & Kuwait & KWT & 1961 - 2016 & Sudan & SDN & 1956 - 2016 \\ 
		Central Afr. Rep. & CAF & 1960 - 2016 & Kyrgyzstan & KGZ & 1991 - 2016 & Suriname & SUR & 1975 - 2016 \\ 
		Chad & TCD & 1960 - 2016 & Laos & LAO & 1950 - 2016 & Swaziland & SWZ & 1968 - 2016 \\ 
		Chile & CHL & 1950 - 2016 & Latvia & LVA & 1991 - 2016 & Sweden & SWE & 1950 - 2016 \\ 
		China & CHN & 1950 - 2016 & Lebanon & LBN & 1950 - 2016 & Switzerland & CHE & 1950 - 2016 \\ 
		Colombia & COL & 1950 - 2016 & Lesotho & LSO & 1966 - 2016 & Syria & SYR & 1950 - 2016 \\ 
		Comoros & COM & 1975 - 2016 & Liberia & LBR & 1950 - 2016 & Taiwan & TWN & 1950 - 2016 \\ 
		DR Congo & ZAR & 1960 - 2016 & Libya & LBY & 1951 - 2016 & Tajikistan & TJK & 1991 - 2016 \\ 
		Congo & COG & 1960 - 2016 & Lithuania & LTU & 1990 - 2016 & Tanzania & TZA & 1961 - 2016 \\ 
		Costa Rica & CRI & 1950 - 2016 & Luxembourg & LUX & 1950 - 2016 & Thailand & THA & 1950 - 2016 \\ 
		Cote dIvoire & CIV & 1960 - 2016 & Macedonia  & MKD & 1991 - 2016 & Timor-Leste & TMP & 2002 - 2016 \\ 
		Croatia & HRV & 1991 - 2016 & Madagascar & MDG & 1960 - 2016 & Togo & TGO & 1960 - 2016 \\ 
		Cuba & CUB & 1950 - 2016 & Malawi & MWI & 1964 - 2016 & Trinidad Tobago & TTO & 1962 - 2016 \\ 
		Cyprus & CYP & 1960 - 2016 & Malaysia & MYS & 1957 - 2016 & Tunisia & TUN & 1956 - 2016 \\ 
		Czech Republic & CZR & 1993 - 2016 & Mali & MLI & 1960 - 2016 & Turkey & TUR & 1950 - 2016 \\ 
		Czechoslovakia & CZE & 1950 - 1991 & Mauritania & MRT & 1960 - 2016 & Turkmenistan & TKM & 1991 - 2016 \\ 
		Denmark & DNK & 1950 - 2016 & Mauritius & MUS & 1968 - 2016 & Uganda & UGA & 1962 - 2016 \\ 
		Djibouti & DJI & 1977 - 2016 & Mexico & MEX & 1950 - 2016 & Ukraine & UKR & 1991 - 2016 \\ 
		Dominican Rep. & DOM & 1950 - 2016 & Moldova & MDA & 1991 - 2016 & Un. Arab Emirates & ARE & 1971 - 2016 \\ 
		Ecuador & ECU & 1950 - 2016 & Mongolia & MNG & 1950 - 2016 & United Kingdom & GBR & 1950 - 2016 \\ 
		Egypt & EGY & 1950 - 2016 & Morocco & MAR & 1956 - 2016 & United States & USA & 1950 - 2016 \\ 
		El Salvador & SLV & 1950 - 2016 & Mozambique & MOZ & 1975 - 2016 & Uruguay & URY & 1950 - 2016 \\ 
		Equatorial Guin. & GNQ & 1968 - 2016 & Myanmar & MMR & 1950 - 2016 & Uzbekistan & UZB & 1991 - 2016 \\ 
		Eritrea & ERI & 1993 - 2016 & Namibia & NAM & 1990 - 2016 & Venezuela & VEN & 1950 - 2016 \\ 
		Estonia & EST & 1991 - 2016 & Nepal & NPL & 1950 - 2016 & Vietnam & VNM & 1976 - 2016 \\ 
		Ethiopia & ETH & 1950 - 2016 & Netherlands & NLD & 1950 - 2016 & South Vietnam & SVM & 1950 - 1975 \\ 
		Fiji & FJI & 1970 - 2016 & New Zealand & NZL & 1950 - 2016 & Yemen & YEM & 1991 - 2016 \\ 
		Finland & FIN & 1950 - 2016 & Nicaragua & NIC & 1950 - 2016 & North Yemen & NYE & 1950 - 1991 \\ 
		France & FRA & 1950 - 2016 & Niger & NER & 1960 - 2016 & South Yemen & SYE & 1950 - 1991 \\ 
		Gabon & GAB & 1960 - 2016 & Nigeria & NGA & 1960 - 2016 & Yugoslavia & YUG & 1950 - 1992 \\ 
		Gambia & GMB & 1965 - 2016 & Norway & NOR & 1950 - 2016 & Zambia & ZMB & 1964 - 2016 \\ 
		Georgia & GEO & 1991 - 2016 & Oman & OMN & 1950 - 2016 & Zimbabwe & ZWE & 1950 - 2016 \\ 
		\hline
	\end{tabular}
	\caption{Countries included in the analysis (columns 1, 4 and 7) with three-digit country codes (columns 2, 5 and 8) and time period of inclusion in the model (columns 3, 6 and 9).}
	\label{count_inc}
\end{table}

\FloatBarrier

\subsection{Details on the estimation procedure}\label{estim_detail}

The recent implementation of Generalised Additive Models (GAM) in the \texttt{R} package \texttt{mgcv} allows for smooth varying coefficients as proposed by \citet{hastie1993}. These models can be represented in GAMs by multiplying the smooths by a covariate (in the given application the smooths of time are multiplied by the covariates. See \citet{wood2017} for more details.

The functions for the smooths are based on P-Splines as proposed by \citet{eilers1996}, giving low rank smoothers using a B-spline basis using a simple difference penalty applied to the parameters. For the smooth time-varying coefficients on the fixed effects a maximum number of 65 knots is used, combined with a second-order P-spline basis (quadratic splines) and a first-order difference penalty on the coefficients.

The non-linear random smooths are estimated similar to those proposed by \citet{durban2005}. As a basic idea, one views the individual smooths as splines with random coefficients, i.e.\ each country has a random effect, that is in fact a function of time that is approximated by regression splines. The parameters of the splines are assumed to be normally distributed with mean zero and the same variance for all curves, which translates into having the same smoothness parameter for all curves. This concept is implemented efficiently in the GAM structure of the \texttt{mgcv} package by using the nesting of the smooth within the respective actor. In order to avoid overfitting and keeping computation tractable, a first-order penalty with nine knots is employed. The smoothness selection is done for all smooths by the restricted maximum likelihood criterion (REML).

As the data set is rather big with  more than 1.3 million observations in the formation model, the fitting procedure of the model is computationally expensive and was virtually impossible with standard implementations in \texttt{R} before the introduction of the \texttt{bam()} function in the \texttt{mgcv}  package in 2016 that needs less memory and is much faster than other comparable packages. The estimation routine employs techniques as proposed in \citet{wood2015}. Those methods use discretization of covariate values and iterative updating schemes that require only subblocks of the model matrix to the computed at once which allows for the application of parallelization tools.

For all computations we also used the statistical programming language \texttt{R} (\citealt{team2014r}). Important packages used for visualization of networks and computation of network statistics are the \texttt{statnet} suite of network analysis packages (\citealt{handcock2008}) as well as the package \texttt{igraph} (\citealt{csardi2006}). For the Tables the \texttt{stargazer} package from \citet{hlavac2013} was employed. For the model evaluation and visualization we used the \texttt{PRROC} package of \citet{grau2015}.

\FloatBarrier

\subsection{Details on the PCA of the time-varying smooth random effects}\label{detail_smooth}
For the analysis of the smooth random effects we are following the discretization approach of \citet[Chap. 8]{ramsay2005}. As noted in Section \ref{heterogen} we assume the random effects $\phi_i(t)$ ($\phi_i^+(t)$ and $\phi_i^-(t)$ in the formation and the persistence model, respectively) to be realizations of a stochastic process $\Phi=\{\phi(t),t\in \tau\}$, for $i=1,...,N$ individual countries and  $\tau=[1951,2016]$.

In order to summarize the information provided by these functions we are searching for a weight function $\beta(t)$ that gives us the principal component scores $\phi_i=\int_{\tau} \beta(t)\phi_i(t) d t$. In order to do so, the weight function $\xi_1(t)$ among all possible functions $\beta(t)$  must be found that maximizes $N^{-1}\sum_{i=1}^{N}(\int_{\tau} \beta(t)\phi_i(t) d t)^2$ subject to the constraint $\int_{\tau} \xi_1^2(t)dt=1$.
From our model we get $N$ individual estimated functions  $\hat{\phi}_i(t)$ for all observations (countries) and can discretize the functions $\hat{\phi}_i(t)$ on a grid. We use $T=100$ equidistant points $\{t_1,...,t_{100}\}$ on the interval $\tau$ of length $|\tau|=\mathcal{T}$. This gives a discretized $(N\times T)$ time series matrix $\hat{\Phi}$ with $N$ country specific observations in the rows and the estimated functions, evaluated at the discrete time points, in the columns:
\begin{equation*}
\hat{\Phi}=\begin{pmatrix} \hat{\phi}_1(t_1) & \cdots & \hat{\phi}_1(t_{100})\\ \vdots & \ddots & \vdots \\ \hat{\phi}_N(t_1) & \cdots& \hat{\phi}_N(t_{100}) \end{pmatrix}
\end{equation*}
Therefore, in fact we are searching for a solution for the discrete approximation of 
\begin{equation*}
\int_{\tau} \beta(t)\hat{\phi}_i(t) d t\approx (\mathcal{T}/T)\sum_{j=1}^{T}\beta(t_j)\hat{\phi}_i(t_j)=\sum_{j=1}^{T}\tilde{\beta}(t_j)\hat{\phi}_i(t_j) 
\end{equation*}
 such that the solution $\tilde{\xi}_1$
 that maximizes the mean square satisfies $||\tilde{\xi}_1 ||^2=1$. This is now a standard problem, with the solution $\tilde{\xi}_1$ being found by the eigenvector that corresponds to the largest eigenvalue of the covariance matrix of $\hat{\Phi}$.

\subsection{Out-of-sample-predictions for simulated networks}\label{osp_sim}

As a standard principle in network analysis, a model should be able the reflect global network characteristics. We evaluate six of them for our out-of-sample forecasts. The first three characteristics are related to the number of actors that are actively engaged in the arms trade. The statistic {\sl Size} is defined as the count of predicted edges in each year. This measure helps to evaluate the ability of the model to predict amount of realized arms trade in each year.	As it is also of interest to measure how dense the predicted arms trade network is, we include {\sl Density}, relating the size of the network to the number of edges that could have potentially realized. We define the {\sl Order} of the network as number of actors that are engaged in either exporting or importing arms. The results will provide an impression whether the model has the ability not only to classify the right amount of edges (as in {\sl Size}), but also their nesting within the countries.

As we have emphasized the importance of local network statistics we evaluate whether the local network statistics are able to generate the corresponding global statistics. Therefore, we include the {\sl Mean Indegree} (being the same as {\sl Mean Outdegree}), as well as the share of  {\sl Reciprocity}. In order to evaluate the accuracy of our predictions with respect to triangular relationships we furthermore include the measure {\sl Transitivity}, that divides the number of triangles by the number of connected triples in the graph. In this statistic, the direction of the edges is ignored. The analysis of this measure gives an impression how well the two chosen transitivity measures capture the overall clustering in the network.

\begin{figure}[t!]
	\centering
	
	\centering\includegraphics[trim={4cm 4.5cm 4cm 4.5cm},clip,width=0.79\textwidth]{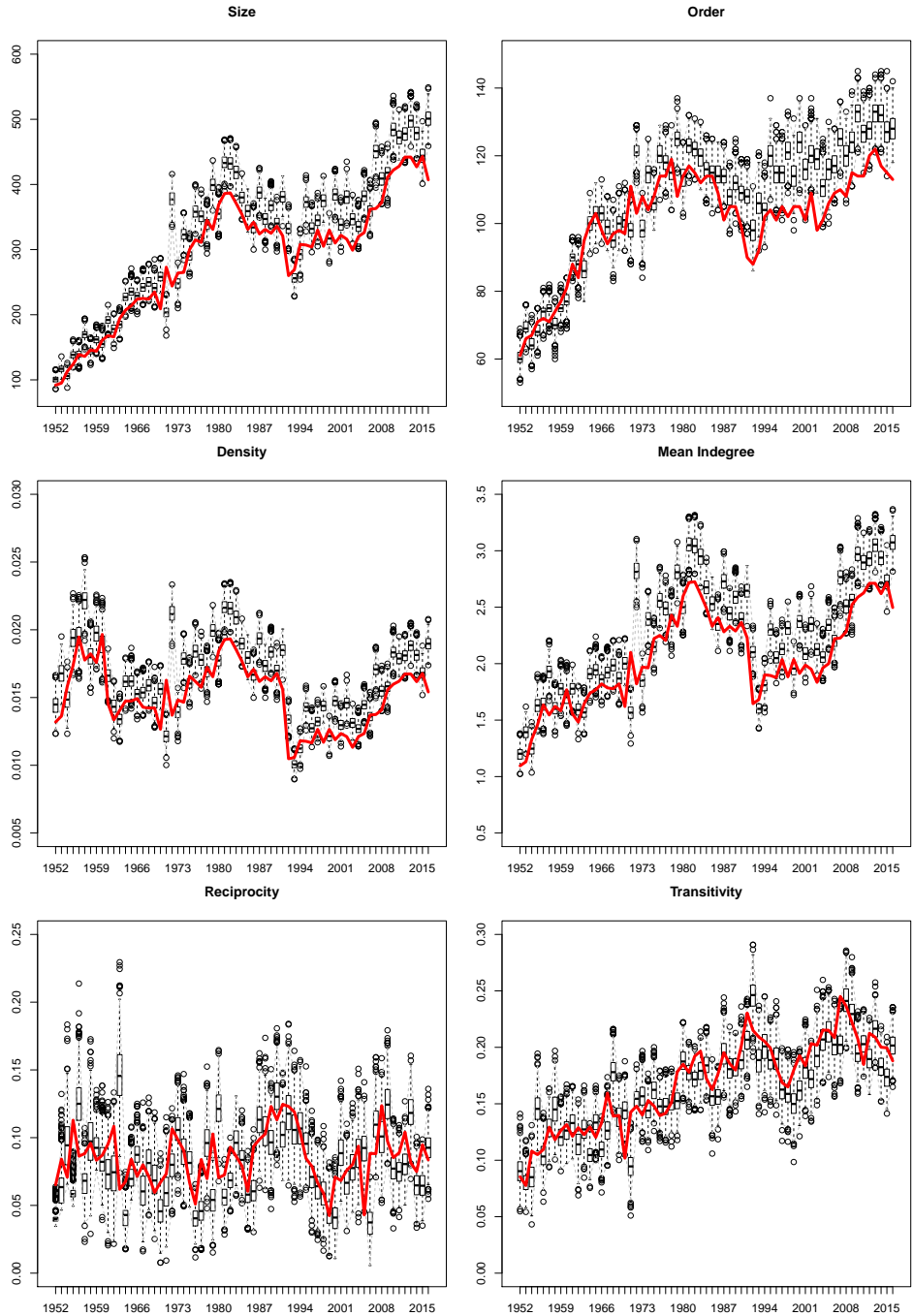}
	
	\caption{Comparison of realized and simulated network topologies.
		The boxplots give statistics from the simulated networks. The solid line  gives the statistics for the real networks. Number of edges (Size), number of active countries (Order), number of realized transfers relative to possible transfers (Density), average indegree (Indegree), share of reciprocated transfers and ratio of undirected triangles relative to connected triples (Transitivity).}
	
	\label{fig_sim1}
\end{figure}

The results are presented in Figure \ref{fig_sim1}. In each of the six panels we see the respective network statistics plotted against time. The solid red line gives the network statistics, evaluated at the real MCW network. The boxplots show the network statistics, evaluated for each year for the 1.000 simulated networks.

\FloatBarrier
\newpage
\section{Supplementary Material}

\subsection{Different threshold values}
\subsubsection{Distribution of TIVs}
\begin{figure}[!t]
	\centering

	\centering\includegraphics[trim={0cm 0cm 0cm 2cm},clip,width=\textwidth]{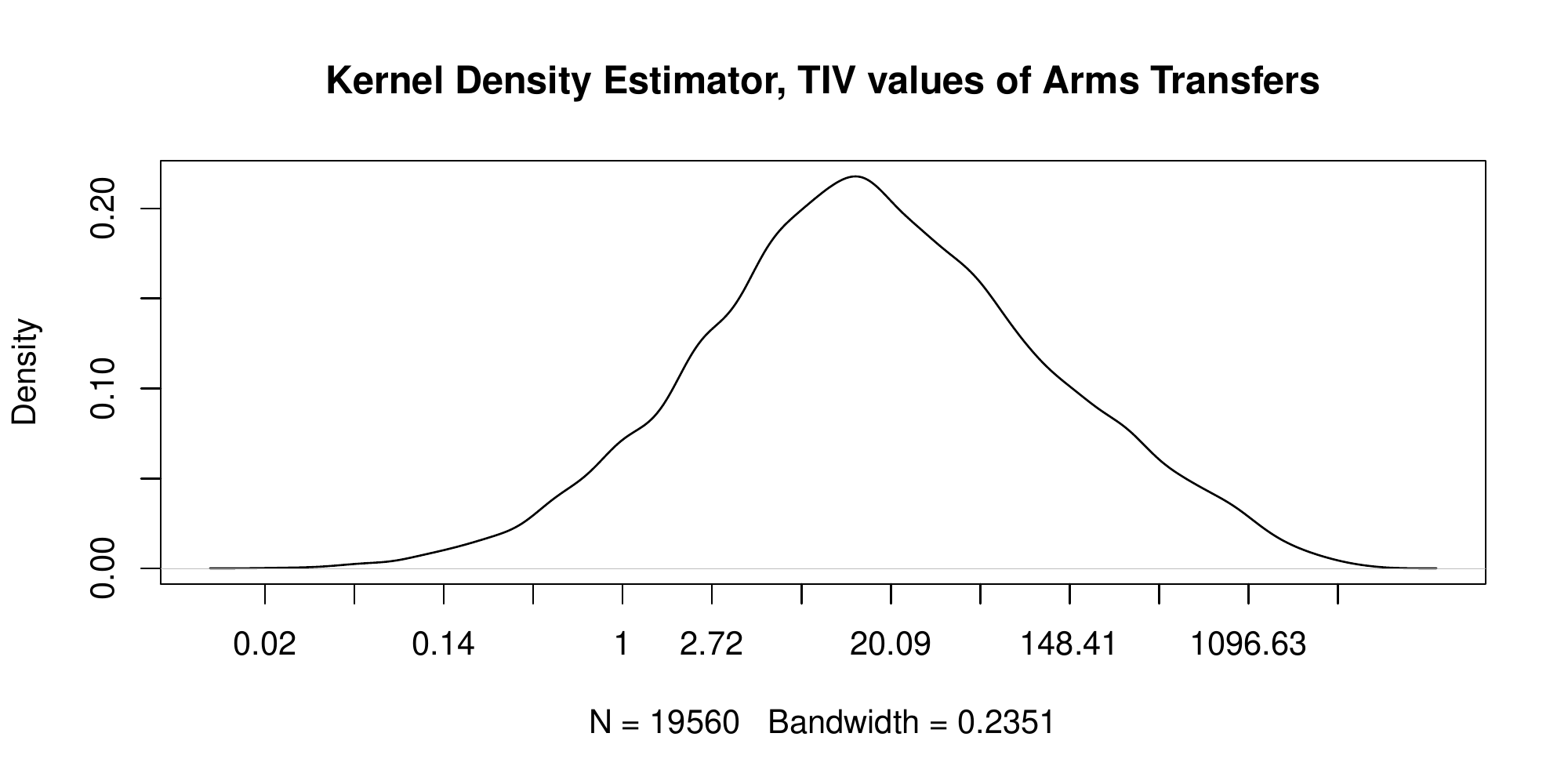}

	\caption{Kernel density estimate (KDE) of arms exports, measured in TIV and pooled over all years from 1950 to 2016. Logarithmic x-axis.}
	\label{fig_denstiy}	
\end{figure}

In Figure \ref{fig_denstiy} we present a kernel density estimate (KDE) of the pooled TIVs for the whole time period. The distribution of the TIVs is highly skewed and has a long tail. Therefore, we give a logarithmic representation. In order to give an impression of the left tail, Table \ref{quantiles} provides the lower quantiles of the distribution. From this it can be seen that roughly 20\% of all observations are below a threshold of $3$.

\begin{table}[!htbp] \centering 
	\begin{tabular}{ccccc} 
		\\ \hline 
		\hline 
		0\% & 4.75\% & 9.5\% & 14.25\% & 19\% \\ 
		\hline 
		$0.020$ & $0.700$ & $1.332$ & $2.200$ & $3$ \\ 
		\hline 
	\end{tabular} 
	\caption{Lower quantiles of the TIV distribution pooled over all years from 1950 to 2016.} 
	\label{quantiles} 
\end{table} 

In order to demonstrate the effect of different binarization thresholds on the estimated coefficients we pursue the following strategy. As a baseline we use the "original effects" from the paper with a threshold of zero and plot them in solid black together with two standard error confidence bounds in dark grey. Additionally, we include the estimated coefficients with thresholds incrementing from zero to three in steps of 0.5 as dashed black lines.  (i) By comparison of the solid line with the dashed lines it can be seen how strong the point-estimates vary with different thresholds. (ii) If the dashed lines are within the confidence intervals in dark grey they can be said to be statistically indistinguishable from the original estimates. (iii) Furthermore, we show how the confidence bounds of the new estimates exceed the ones of the original estimation, displayed in light grey. These areas represent the highest upper limit and the lowest lower limit that exceeds the original confidence bounds.  Hence, if only the dark grey confidence bound is visible, then the effects of all estimates are the same, if the light grey confidence bound is above and/or below the dark grey, this means that the bounds of the estimates with higher thresholds are wider. 
\subsubsection{Fixed effects with different thresholds}
\begin{figure}[!t]
	\centering

	\centering\includegraphics[trim={0cm 0cm 0cm 0cm},clip,width=0.9\textwidth]{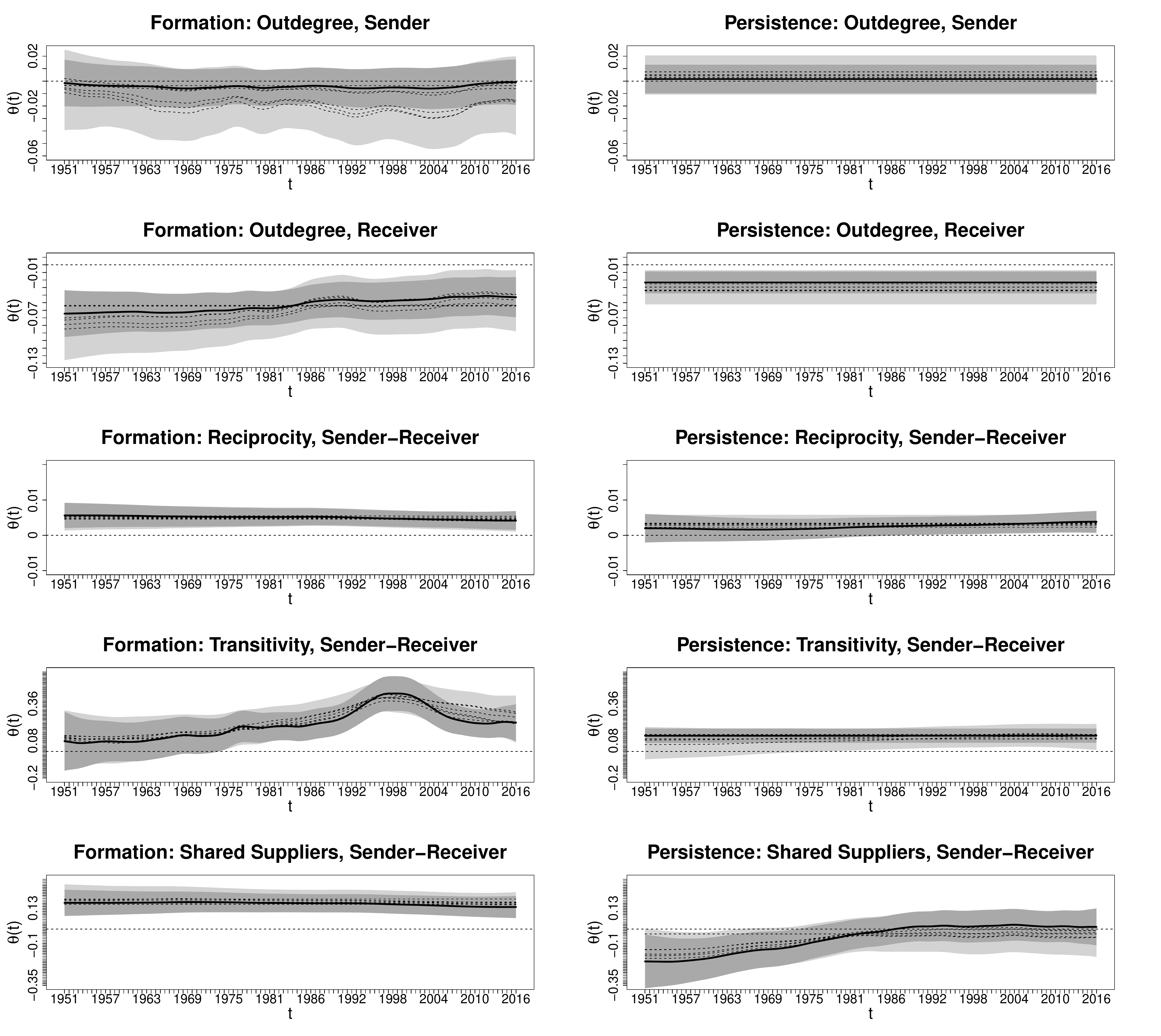}

	\caption{Comparison of fixed effects for network statistics with different binarization thresholds ranging from zero to three, incrementing by 0.5. The estimates with a threshold of zero are given in solid black with dark grey confidence bounds. All other estimates are indicated by dashed lines. Confidence bounds derived from estimates with higher thresholds in light grey.}
	\label{fig_comp1}	
\end{figure}
The degree-related statistics are shown in the top four panels of Figures \ref{fig_comp1}. We find no significant changes of the results as the dashed lines stay in almost all cases within the original confidence bounds. An exception is the senders' outdegree in the formation model but even with the highest threshold this effect does not become significant. On the contrary, partly the results get even more clear. For example the outdegree effect for the receiver in the formation and dissolution model (second row) becomes even more negative in tendency with increasing thresholds.

For the reciprocity effect in the third row we find that the coefficients stay almost the same for all different thresholds of binarization. The same applies for the transitivity effect in the formation model (left panel in the fourth row). Here the upper confidence bound even indicates a potentially higher effect. For the transitivity in the persistence model (right panel in the fourth row) we find that the effect becomes insignificant in the beginning if we set roughly 20\% of the lowest observations to zero. Otherwise the effect stays significant and very close to the point estimates of the original estimation.

For the Shared Suppliers Effect (the two panels at the bottom) the results are very similar to the transitivity effects, i.e. we find an potentially stronger effect in the formation model (left panel at the bottom) and an insignificant effect for the first years in the persistence model (right panel) with the highest binarization threshold.

Given that the network statistics are constructed from the network and therefore directly and potentially strongly affected by different thresholds, the robustness of the effects is reassuring. Only if we replace almost one fifth of the existing edges by zeros the effects of the hypderdyadic statistics in the persistence model start to become partly insignificant. However, the affected statistics are in line with our theoretical expectation that the network effects matter mostly for the formation.

\begin{figure}[!t]
	\centering

	\centering\includegraphics[trim={0cm 0cm 0cm 0cm},clip,width=0.9\textwidth]{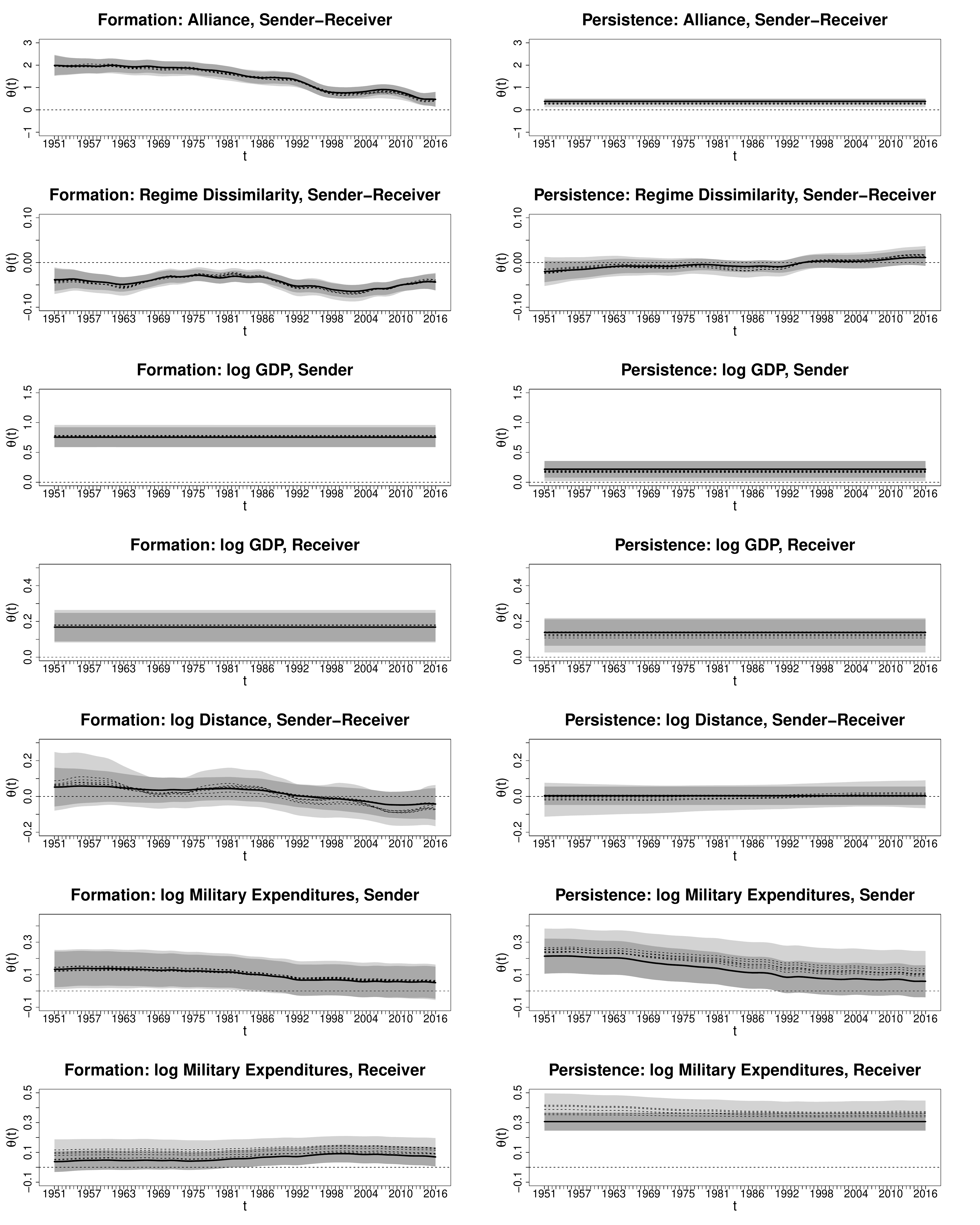}

	\caption{Comparison of fixed effects for economic and political Covariates with different binarization thresholds ranging from zero to three, incrementing by 0.5. The estimates with a threshold of zero are given in solid black with dark grey confidence bounds. All other estimates are indicated by dashed lines. Confidence bounds derived from estimates with higher thresholds in light grey.}
	\label{fig_comp2}	
\end{figure}

From this result it might not come as a surprise that the effects of the non-network related covariates are even more robust because their construction is not affected by the thinning of the network. We can confirm \citet{akerman2014} with Figure \ref{fig_comp2} that shows  virtually no noteworthy changes of the effects.
\FloatBarrier
\subsection{Different time windows}
In the paper we assume that the STERGM process applies to two consecutive years. As a robustness check, we define the periods $t$ and $t+1$ such that they contain multiple years. If we take the years 2013, 2014, 2015 and 2016 as an example for time windows of length two, we set $Y^{t-1,t}_{ij}=1$ if country $i$ exports arms to $j$ in 2013 or 2014 and   $Y^{t,t-1}_{ij}=0$ if country $i$ has not exported to $j$ neither in 2015 nor in 2016. We also extend this concept such that we combine three years into one period. 

For the non-network related covariates we are using the time averages for the respective time windows for continuous variables (e.g.\ if a period contains two years, the average of the GDP in these two years is taken) and we set binary variables to one if the respective feature was present in all year (e.g.\ the indicator for a formal alliance is one if the alliance was present in all two or three years). 

\begin{figure}[!t]
	\centering

	\centering\includegraphics[trim={0cm 0cm 0cm 0cm},clip,width=0.9\textwidth]{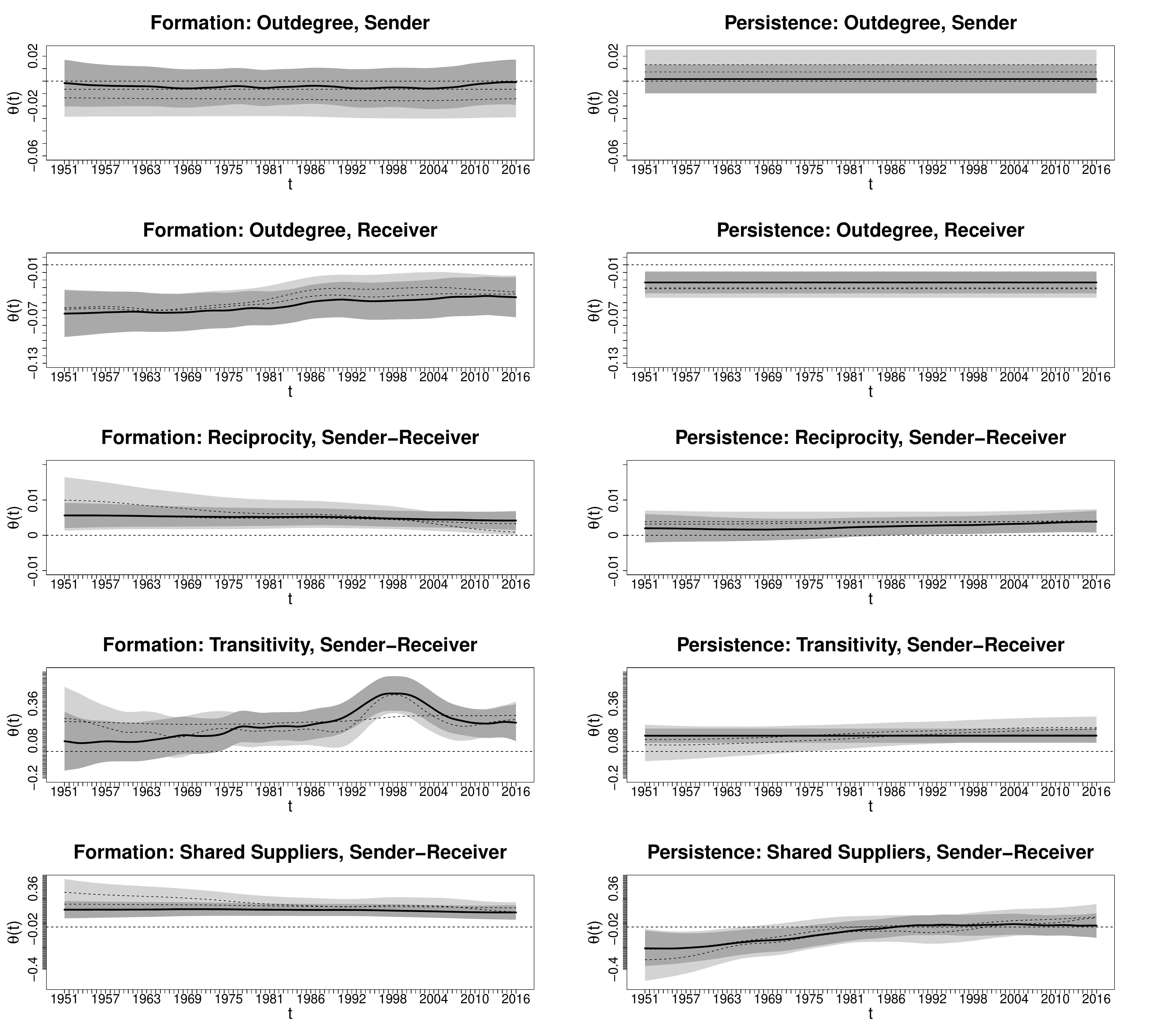}

	\caption{Comparison of fixed effects for network statistics with different time windows. The original estimates are given in solid black with dark grey confidence bounds. The estimates with time windows containing two years are given in dotted and those with three years in dotted-dashed. Confidence bounds derived from estimates with broader time windows in light grey.}
	\label{fig_window1}	
\end{figure}

\begin{figure}[!t]
	\centering

	\centering\includegraphics[trim={0cm 0cm 0cm 0cm},clip,width=0.9\textwidth]{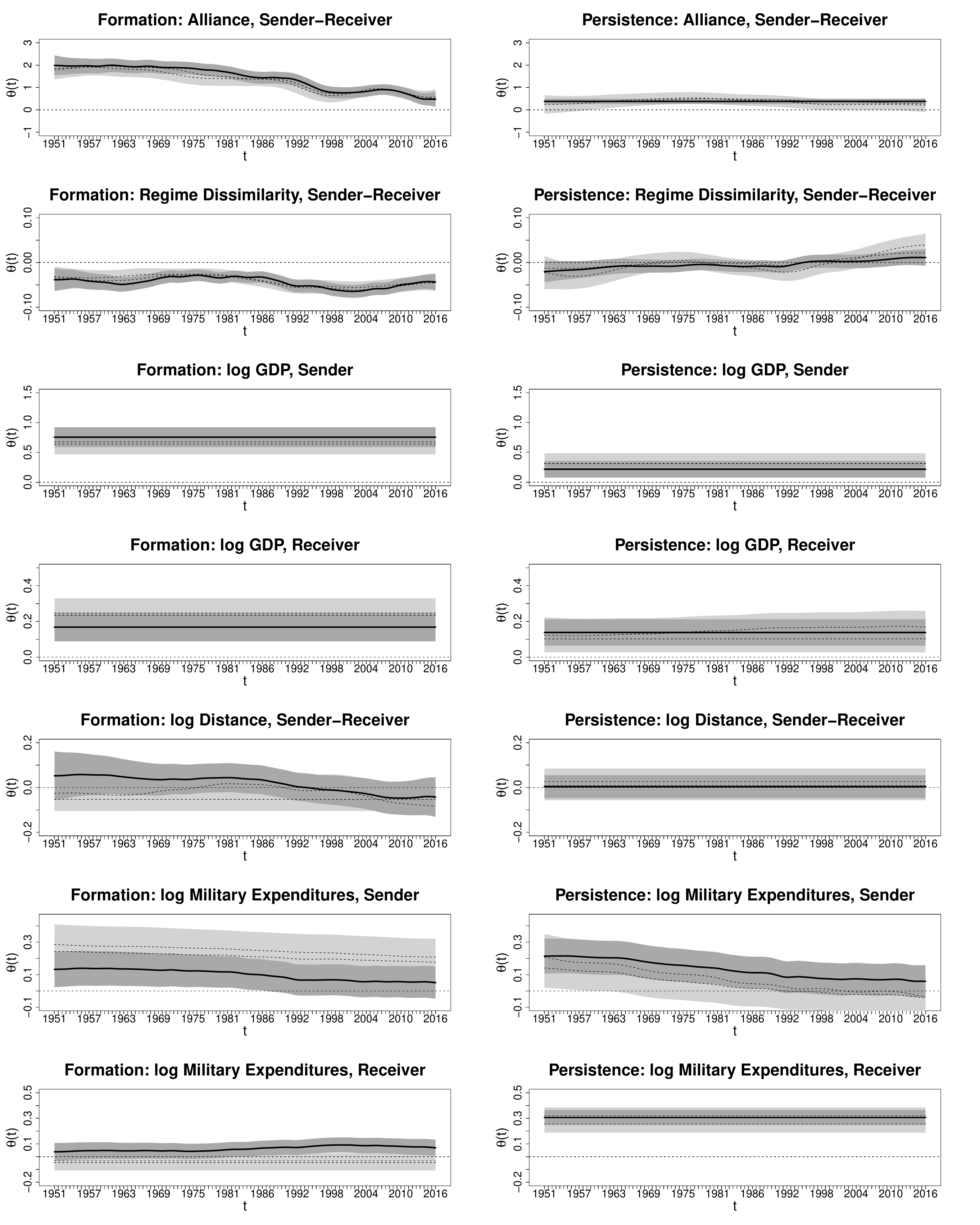}

	\caption{
		Comparison of fixed effects for economic and political covariates with different time windows. The original estimates are given in solid black with dark grey confidence bounds. The estimates with time windows containing two years are given in dotted and those with three years in dotted-dashed. Confidence bounds derived from estimates with broader time windows in light grey.}
	\label{fig_window2}	
\end{figure}

The corresponding estimates can be seen in Figures \ref{fig_window1} and \ref{fig_window2}. (i) The Figures are constructed such that the baseline is given by the "original effects" from the paper. These effects are plotted in solid black together with two standard error confidence bounds in dark grey. (ii) The coefficients with two or three years within one time period are given in dotted (two years) and dotted-dashed (three years). (iii) The area where the original  confidence bounds are exceeded is given in light grey.

In Figure \ref{fig_window1} it can be seen that the dotted and dashed-dotted line rarely is outside of the dark grey confidence bound in the panels of rows one to three. One really noteworthy exception applies to the transitivity effect in the formation model. Here the estimates with periods containing three years behave somewhat more time-constant than the original estimates. This is, however, a natural result because with the broader time windows major changes in the data as the collapse of the Soviet Union become more smooth. All in all, the panels clearly show that the coefficients are very robust and do not change fundamentally. This impression is confirmed by  Figure \ref{fig_window2} where again no notable exceptions can be found influence the interpretation in terms of variation with time or significance. This is again in line with our theoretical expectations since the coefficient for the military expenditures of the receiver is almost not affected by different time windows.

\FloatBarrier
\subsection{Model without random effects}
\begin{figure}[!t]
	\centering

	\centering\includegraphics[trim={0cm 0cm 0cm 0cm},clip,width=0.9\textwidth]{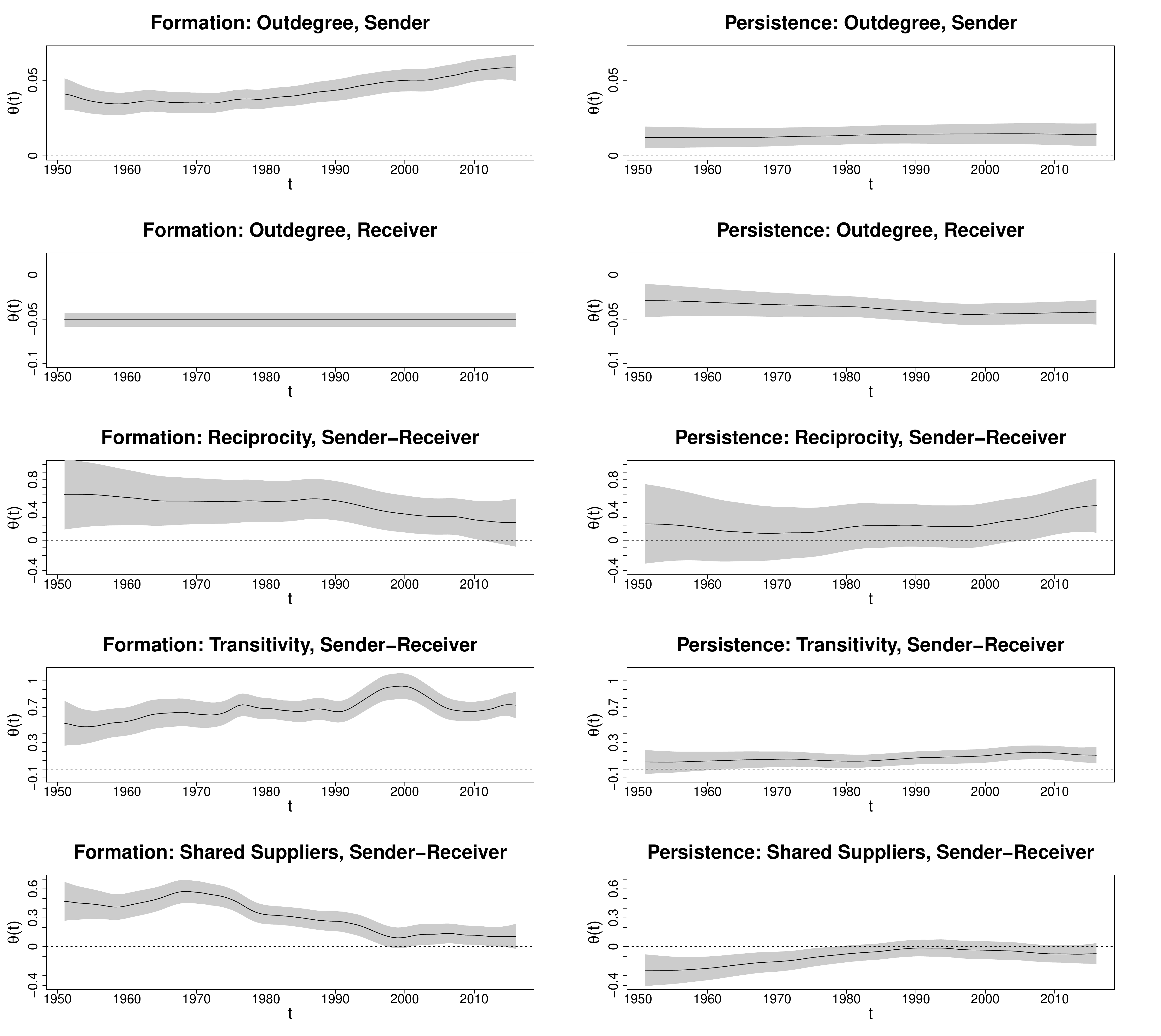}

	\caption{Time-varying coefficients of network statistics without random effects in solid black. Shaded areas give two standard error bounds.}
	\label{fig_no_raneff1}	
\end{figure}

\begin{figure}[!t]
	\centering

	\centering\includegraphics[trim={0cm 0cm 0cm 0cm},clip,width=0.9\textwidth]{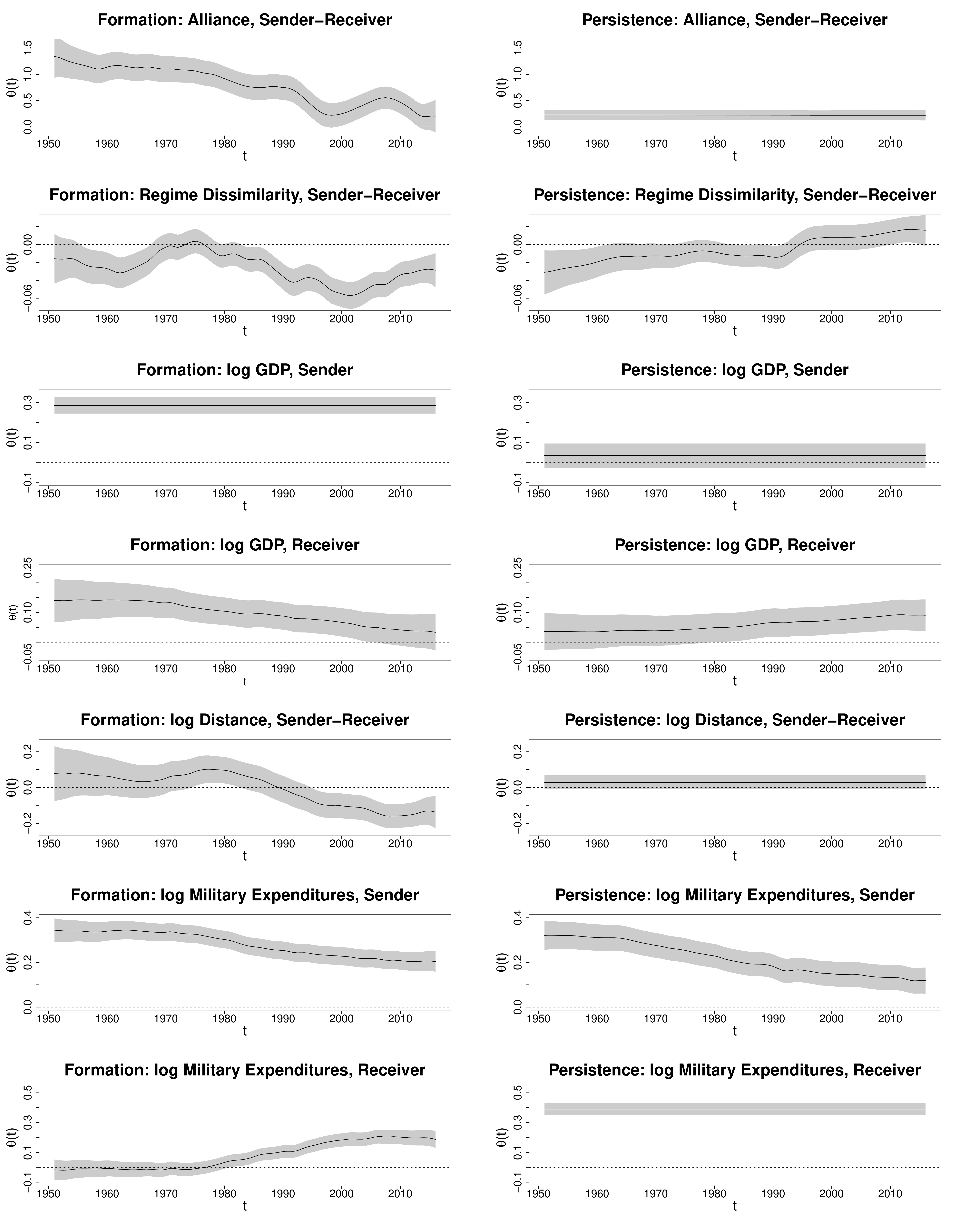}

	\caption{Time-varying coefficients of economic and political covariates without random effects in solid black. Shaded areas give two standard error bounds.}
	\label{fig_no_raneff2}	
\end{figure}

In the main article we mentioned that the inclusion of the random effects leads to a vanishing global effect of the senders outdegree. I.e\ once controlled for the sender-specific random effect the coefficient on the outdegree statistic is insignificant. Here, we show that once we exclude random effects from the models, all results are very robust with the exception of the coefficient on the outdegree. This can be seen in Figures \ref{fig_no_raneff1} and \ref{fig_no_raneff2} with coefficients that are very comparable to the ones from the main paper. The main exception is given by the senders outdegree (top panels in Figure \ref{fig_no_raneff1}). Here the coefficients are now in both models positive and significant. Especially in the formation effect the coefficient is very high. This shows that there is indeed a global effect of the senders outdegree that, however, vanishes if one controls for country-specific heterogeneity.

\FloatBarrier
\subsection{Comparison of different methods}

\subsubsection{Theoretical discussion}
Besides the STERGM, there are many other models suitable for dynamic networks. In the following, we provide an overview of alternative approaches with a discussion of their suitability for the given dataset.

\noindent \textsl{Latent Space Models:} A potential alternative is given by the family of latent space models (e.g.\ \citealt{hoff2002}; \citealt{handcock2007}). However, it is hard to model arms trade appropriately within this model class. The need of estimating time-varying coefficients results in yearly separate estimations. The resulting yearly latent space representations are not very stable and do not work well with isolates, i.e.\ countries without transfers in a given year. If one goes for a panel approach with this model class (see the package \texttt{amen} by \citealt{amen2015}) one must accept that the coefficients stay the same for the whole time-period and, even more problematic, that the positions within one latent space are sufficient to capture all network dependencies for the time-period 1950-2016, which is clearly a heroic assumption.
On top of that, the latent space approach does not allow for the evaluation of complex network statistics we are interested in.

\noindent \textsl{Stochastic-Actor Oriented Models:} Stochastic-Actor Oriented Model (SAOM, see e.g. \citealt{snijders2010soam}) are built for modelling dynamic networks and have the virtue of allowing for the estimation of dyadic and hyper-dyadic network effects. However, the model is tailored for social networks and some of its' assumptions are very problematic for modelling the arms trade network. Most importantly, it assumes actor-homogeneity, an assumption that is clearly violated for the dataset. Additionally, the SAOM fundamentally builds on the idea that the networks observed represent snapshots of a continuous underlying process of edge formation and persistence, i.e.\ we would need to assume that between $t$ and $t+1$ multiple changes could have been realized in the network. This is not an acceptable assumption because if a transfer between $i$ and $j$ was recorded in $t$ and $t+1$, it is not meaningful to assume that there exists an in-between state where the transfer has the change to disappear and re-emerged multiple times. Otherwise, if no transfer is recorded in $t$ and $t+1$ we have no reason to believe that there was trade in between.
In the basic description of the SAOM (\citealt[p. 54]{snijders2010soam})  the authors write: \textit{"A foundational assumption of the models discussed in
	this paper is that the network ties are not brief events, but can be
	regarded as states with a tendency to endure over time. Many relations
	commonly studied in network analysis naturally satisfy this
	requirement of gradual change, such as friendship, trust, and cooperation."} Apparently it is hard to argue that recording whether there was a transfer between two countries in a given year can be viewed as an enduring state. 

\noindent \textsl{Exponential Random Graph Models:} 
Our model is in fact motivated by recent advances within the exponential random graph model (ERGM) family, i.e.\ the TERGM (\citealt{hanneke2010}, \citealt{snijders2010}, \citealt{leifeld2017}) and the STERGM (\citealt{krivitsky2014}).
However, our model differs from the standard cross-sectional ERGM (but also from the conventional TERGM and STERGM) because it does not allow for simultaneous network dependencies. In general, a static cross-sectional ERGM seems to be an implausible choice for the modelling of a dynamic network with strong actor heterogeneity. A dynamic formation and the TERGM, however, can be constructed in a very similar manner as the STERGM. Those models are natural candidate model for comparison.

\subsubsection{Candidate models}
In the following we present several alternative candidate models by increasing level of complexity.

\noindent \textsl{Autoregressive ERGM (Model 1):}
The most simplistic stochastic model is an autoregressive model that assumes time-dependence of all individual dyads such that
\begin{equation}
\label{logit1}
\log\bigg{\{}  \frac{P(Y^{t,t-1}_{ij}=1|Y^{t-1,t}=y^{t-1,t},X^{t-1,t}=x^{t-1,t};\theta)}{P(Y^{t,t-1}_{ij}=0|Y^{t-1,t}=y^{t-1,t},X^{t-1,t}=x^{t-1,t};\theta)}\bigg{\}}=\theta_0+\theta_1y_{ij}^{t-1,t}.
\end{equation}
This temporal dependence structure can be interpreted as a cross sectional ERGM with the lagged response as a dyadic exogenous covariate or as a TERGM with a dyadic stability term (see e.g.\ \citealt{block2018}). It is motivated by the idea that the probability of a transfer in $t$ might change if there was a transfer in $t-1$. 

\noindent \textsl{TERGM with covariates (Model 2):} We can change model (\ref{logit1}) by including all the network effects and covariates as specified in Section 3.2 of the main paper:
\begin{equation}
\label{logit2}
\log\bigg{\{}  \frac{P(Y^{t,t-1}_{ij}=1|Y^{t-1,t}=y^{t-1,t},X^{t-1,t}=x^{t-1,t};\theta)}{P(Y^{t,t-1}_{ij}=0|Y^{t-1,t}=y^{t-1,t},X^{t-1,t}=x^{t-1,t};\theta)}\bigg{\}}=\theta \tilde{g}_{ij}(y^{t-1,t},x^{t-1,t})
\end{equation}
In this formulation, we include the autoregressive component only indirectly, the lagged network statistics give some information about the network embedding of a transfer but not  whether there was a preceding transfer.

\noindent \textsl{TERGM with covariates and random effects (Model 3):}
With the inclusion of smooth time-varying random effects for the sender and the receiver we have
\begin{equation}
\begin{split}
\label{logit3}
\log\bigg{\{}  \frac{P(Y^{t,t-1}_{ij}=1|Y^{t-1,t}=y^{t-1,t},X^{t-1,t}=x^{t-1,t};\theta)}{P(Y^{t,t-1}_{ij}=0|Y^{t-1,t}=y^{t-1,t},X^{t-1,t}=x^{t-1,t};\theta)}\bigg{\}}=&\theta \tilde{g}_{ij}(y^{t-1,t},x^{t-1,t})\\&+\phi_{i,sender}(t)+\phi_{j,receiver}(t).
\end{split}
\end{equation}
The main difference to the STERGM is now that we do not model the processes of formation and persistence separately but within one model and that we do not include the information on the lagged response here, that is implicitly included in the STERGM mechanics.

\noindent \textsl{TERGM with covariates and dyadic stability (Model 4):} The inclusion of the lagged response to the TERGM with covariates gives
\begin{equation}
\label{logit4}
\log\bigg{\{}  \frac{P(Y^{t,t-1}_{ij}=1|Y^{t-1,t}=y^{t-1,t},X^{t-1,t}=x^{t-1,t};\theta)}{P(Y^{t,t-1}_{ij}=0|Y^{t-1,t}=y^{t-1,t},X^{t-1,t}=x^{t-1,t};\theta)}\bigg{\}}=\theta \tilde{g}_{ij}(y^{t-1,t},x^{t-1,t})+\theta_1y_{ij}^{t-1,t}.
\end{equation}

\noindent \textsl{TERGM with network effects, dyadic stability and random effects (Model 5):} And as a last step we allow for sender- and receiver-specific random effects in the TERGM with network statistics and lagged response:
\begin{equation}
\label{logit5}
\begin{split}
\log\bigg{\{}  \frac{P(Y^{t,t-1}_{ij}=1|Y^{t-1,t}=y^{t-1,t},X^{t-1,t}= x^{t-1,t};\theta)}{P(Y^{t,t-1}_{ij}=0|Y^{t-1,t}=y^{t-1,t},X^{t-1,t}=x^{t-1,t};\theta)}\bigg{\}}=&\theta \tilde{g}_{ij}(y^{t-1,t},x^{t-1,t})+\theta_1y_{ij}^{t-1,t}\\&+\phi_{i,sender}(t)+\phi_{j,receiver}(t).
\end{split}
\end{equation}
\noindent \textsl{STERGM without random effects (Model 6):}  
Together with our main model from the paper (Model 7), the STERGM with random effects we additionally include a STERGM without random effects. The formal representations are given in the main paper in equations (4) and (5).
\begin{table}[t!]\small
	\begin{tabular}{llccccc}\hline
		Number & Model Name                                                   & lagged edge & covariates & rand. eff. & PR \% & ROC \%\\ \hline \hline
		1            & Autoreg. ERGM                                         & yes         & no              & no      &0&0       \\
		2            & TERGM                             & no          & yes             & no          &0&0    \\
		3            & TERGM            & no          & yes             & yes        &0&1.54\%         \\
		4            & TERGM with  dyadic stability               & yes         & yes             & no        &4.62\%&6.15\%      \\
		5            & TERGM with dyadic stability  & yes         & yes             & yes        &18.46\%&44.62\%         \\
		6            & STERGM                                                       & implicit    & yes             & no   &24.62\%&6.15\%           \\
		7            & STERGM                               & implicit    & yes             & yes           &52.31\%&41.54\%  
		\\ \hline
	\end{tabular}
	\caption{Different dynamic network models from the ERGM family included in the comparison of prediction. Name of the models in the second column. Model specification in columns three, four and five. The share of years where the respective model performed best according to the AUC of the PR or the ROC curve are given in the two rightmost columns.} \label{models}
\end{table}
\FloatBarrier
\subsubsection{Comparison of out-of-sample predictions}
We evaluate the proposed models in the following way. In a first step, we fit coefficients based on the information of  $t-1$ to the response in  $t$. In the second step we use the fitted models in order to predict the edges in $t+1$. As a result we obtain probabilistic out-of-sample predictions. 
The evaluation is done with area under the curve (AUC) measures for the Receiver-Operating-Characteristic (ROC) curve and the Precision-Recall (PR) curve.

All models are fitted using the package \texttt{mgcv} (\citealt{wood2017}; Version 1.8-24) and evaluated with the package \texttt{PRROC} (\citealt{grau2015}; Version 1.3).

Table \ref{models} gives and overview of the predictive performance of all models included. In the two rightmost columns we present the share of years where the respective model has the highest out-of-sample predictive power. It can be seen that evaluated by the PR, the STERGM with random effects clearly represents the superior model with the highest predictive performance. Judged by the ROC the performance of the TERGM and the STERGM (both with random effects and as a model class) are very similar. We give a detailed description of the results below.

In order to give a clear impression of how the out-of-sample fits are related we provide multiple plots where the baseline model, the STERGM with covariates and random effects, is always indicated in solid red.
\begin{figure}[!t]
	\centering

	\centering\includegraphics[trim={0cm 0cm 0cm 0cm},clip,width=\textwidth]{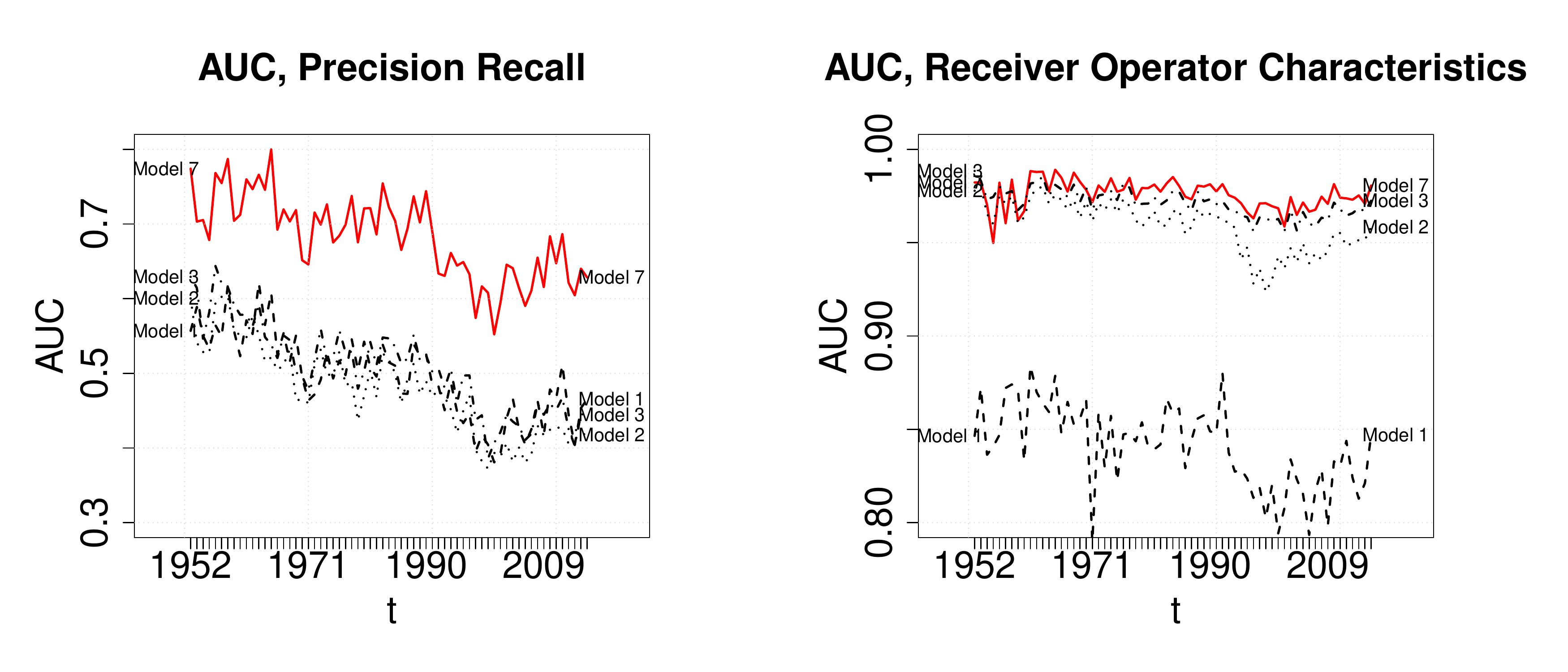}

	\caption{AUC values for out-of-sample predictions based on precision recall (left) and receiver-operator-characteristics (right). The STERGM with covariates and random effects (Model 7) in solid and red, the autoregressive ERGM (Model 1) and the TERGM with covariates (Model 2) and random effects (Model 3) in black and dashed.}
	\label{fig_roc1}	
\end{figure}

In Figure \ref{fig_roc1} we compare the AUC values of the autoregressive ERGM (Model 1), the TERGM with covariates (Model 2) and with random effects (Model 3)
with the baseline STERGM model (Model 7). The Precision Recall AUC values are shown on the left hand side of  Figure \ref{fig_roc1} and provide a clear message since all selected candidate models have AUC values clearly below the STERGM used in the paper (Model 7).  Looking at the AUC values from the ROC measure (right panel) shows that the two TERGM models (Models 2 and 3) come partly close to the baseline model in the first years of the observational period, while the simplistic autoregressive ERGM has considerable lower AUC values. However as a general picture the STERGM with random effects is clearly the superior model.
\FloatBarrier
Including the lagged response as explanatory variable makes the TERGM models (Models 4 and 5) pretty similar to our baseline model (Model 7).
\begin{figure}[!t]
	\centering
	\centering\includegraphics[trim={0cm 0cm 0cm 0cm},clip,width=\textwidth]{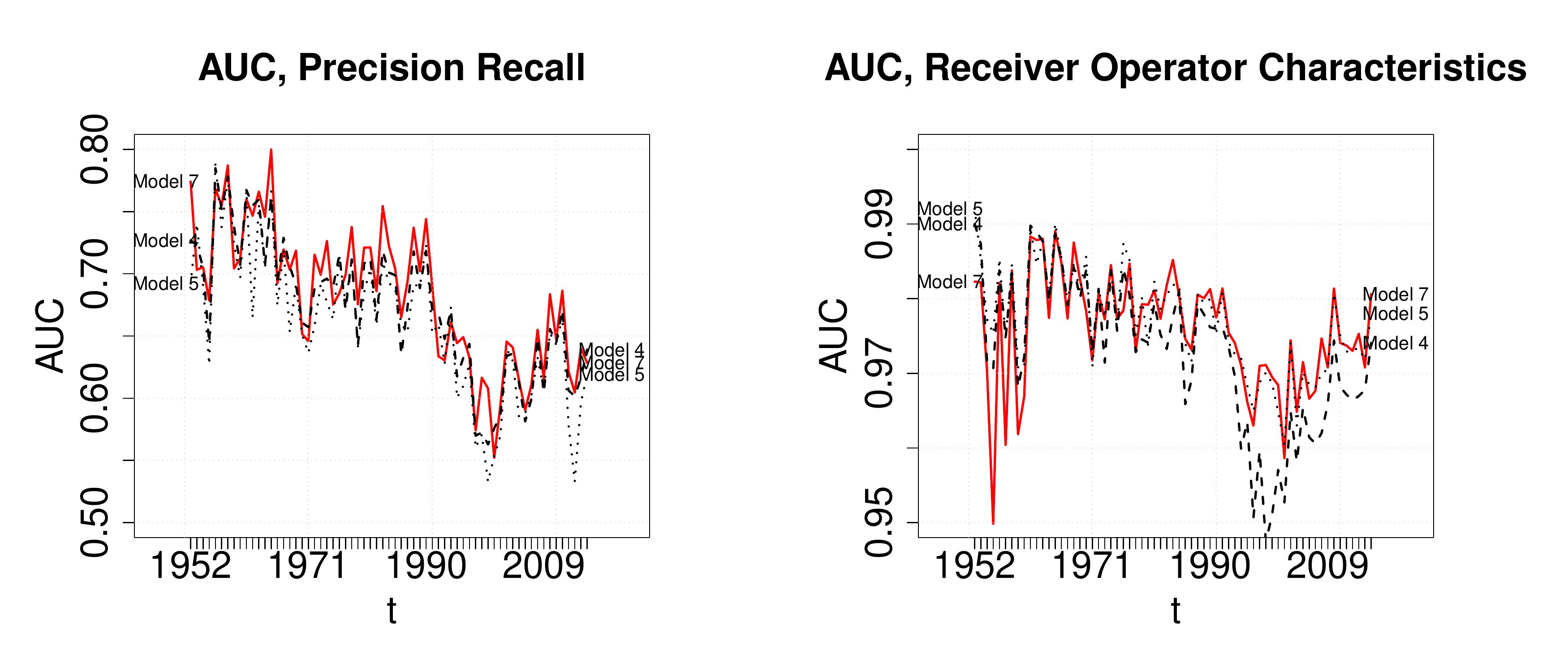}
	\caption{AUC Values for out-of-sample predictions based on precision recall (left) and receiver-operator-characteristics (right). The STERGM with covariates and random effects (Model 7) in solid and red, the TERGM with covariates, dyadic stability (Model 4) and random effects (Model 5) in black and dashed.}
	\label{fig_roc2}	
\end{figure}
Naturally, this is reflected in Figure \ref{fig_roc2}. However, the STERGM model is more flexible because it allows for different coefficients for the processes of formation and persistence. This fact leads to superior predictions of the STERGM model (Model 7) in the left panel of Figure \ref{fig_roc2}. The red line provides the upper boundary in most time points, but there are some instances where the autoregressive TERGMs (Models 4 and 5) provide the better predictions when evaluated with the PR curve. If the AUC measure for the ROC curve is compared, we see again that the predictions of the TERGM partly outperform the STERGM.  Nevertheless, this is mostly the case in the beginning of the observational period and it seems like the superiority of the STERGM increases with the more recent periods. 

\begin{figure}[!t]
	\centering
	
	\centering\includegraphics[trim={0cm 0cm 0cm 0cm},clip,width=\textwidth]{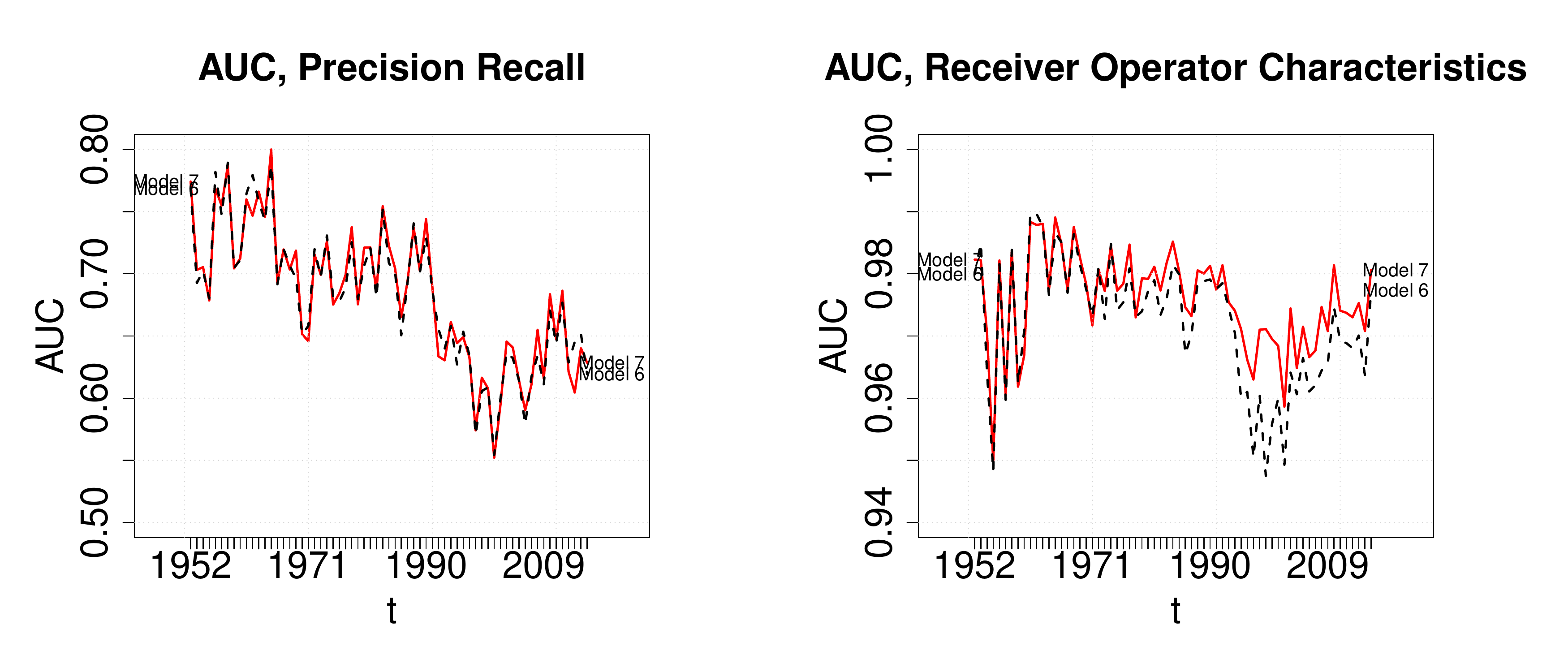}

	\caption{AUC values for out-of-sample predictions based on precision recall (left) and receiver-operator-characteristics (right). The STERGM with covariates and random effects (Model 7) in solid and red, the STERGM with covariates but without random effects (Model 6) in black and dashed.}
	\label{fig_roc3}	
\end{figure}

In Figure \ref{fig_roc3} the predictive performance of the STERGM with (Model 7) and without (Model 6) random effects is compared. Although the AUC values are very close to each other, the model that includes the random effects clearly provides better out-of-sample predictions. On the right hand side of Figure \ref{fig_roc3}  the contrast becomes visible more clearly and in almost all years our baseline model provides the better predictions.

All in all we conclude the following. (i) The STERGM gives, among all other candidate models the best predictions when judged by Precision Recall, being the more important measure when predicting rare vents. (ii) Furthermore, the STERGM has a much richer interpretation than the TERGM and (iii) the random effects provide an substantial benefit to the inferential part of the model. We therefore conclude that the choice of the STERGM with random effects seems to be very appropriate regarding both, the predictive performance as well as the ability to gain new insights.
\FloatBarrier

\end{document}